\documentclass[aps,prd,showpacs,preprintnumbers]{revtex4-2}
\usepackage{amsmath}
\usepackage[colorlinks=true,linktocpage=true,linkcolor=blue,citecolor=blue]{hyperref}
\usepackage{subfig}
\usepackage{graphicx}
\usepackage{epstopdf}
\usepackage{float}
\usepackage{subfloat}
\usepackage{amssymb,amsfonts}
\usepackage{latexsym}
\usepackage{epsfig}
\usepackage{amssymb}
\usepackage{color}
\usepackage{bm}
\usepackage{ulem}
\newcommand{\eq}{\begin{equation}}
\newcommand{\eeq}{\end{equation}}
\newcommand{\eqn}{\begin{eqnarray}}
\newcommand{\eeqn}{\end{eqnarray}}

\begin{document}
\title{Production rate and ellipticity of lepton pairs from a rotating hot and dense QCD medium}
\author{Minghua Wei$^{1,2}$}
\thanks{weimh@ihep.ac.cn}
\author{Chowdhury Aminul Islam$^{1}$ }
\thanks{chowdhury.aminulislam@ucas.ac.cn} 
\author{Mei Huang$^{1}$}
\thanks{huangmei@ucas.ac.cn}
\affiliation{$^{1}$ School of Nuclear Science and Technology, University of Chinese Academy of Sciences, Beijing 100049, China}
\affiliation{$^{1}$ Institute of High Energy Physics, Chinese Academy of Sciences,Beijing 100049, China}

\begin{abstract}
Using a current-current correlation function (CF), the photon polarization tensor is calculated for a rotating hot and dense QCD medium. The spectral function (SF) and the dilepton rate (DR) are estimated therefrom.  Numerical results show that both SF and DR are enhanced in a rotating medium, especially in a low invariant mass region. SF and DR are also explored in the consequences of the interplay among the angular velocity, temperature and chemical potential. We also estimated the electromagnetic screening by calculating the Debye mass and it shows a suppression for a rotating QCD medium. The most interesting observation is the azimuthal anisotropy of the dilepton production, i.e, the elliptic flow $v_{2}$ of the lepton pair induced by the rotation as an external field. The competition between the centrifugal effect and the spin polarization effect due to rotation results in a convex down behavior of the elliptic flow as a function of the transverse momentum in a relatively large magnitude of angular velocity. It is noticed that quark spin polarization induces a negative $v_2$ in the case of large angular velocity.
\end{abstract}

\maketitle

\section{Introduction}

It is expected that the deconfined state of quark-gluon plasma (QGP) can be created through heavy-ion collisions (HIC)~\cite{Muller:1983ed,Heinz:2000bk}. The QGP at high temperature is supposed to be present after a few microseconds of the big bang in the early universe and the deconfined dense QCD matter can exist in the core of neutron star~\cite{Yagi:2005yb}. Thus the study and characterization of such quantum chromodynamics (QCD) medium are of great importance.

The QCD medium created in HICs or present in nature can be subjected to different extreme environments, e.g.,  high temperature and/or density, strong magnetic field and rapid rotation, etc ~\cite{Fukushima:2011jc,Kharzeev:2013jha,Chen:2021aiq}. 
It is obvious that the properties of the created medium will be affected by such external conditions, which attracts much interest from the theorists to investigate such a system in detail.

Particularly, in the non-central heavy-ion collision(HIC), large vorticity and strong magnetic field will be generated in the early stage of quark-gluon plasma(QGP). Straightforward computation shows that the magnitude of the magnetic field would reach $10^{18-19}$ Gauss \cite{Kharzeev:2007jp,Skokov:2009qp}, and simulations from kinetic theory and hydrodynamics \cite{Becattini:2007sr,Jiang:2016woz} indicate that the local vorticity would exceed $0.5 fm^{-1}$ with the total angular momentum of QGP at a range of $10^{4}-10^{5}\hbar$.  With the presence of a background magnetic field, the strongly interacting matter shows some nontrivial phenomena, for example, Chiral Magnetic Effect (CME)~\cite{Kharzeev:2007tn,Kharzeev:2007jp,Fukushima:2008xe,Kharzeev:2010gr}, Magnetic Catalysis (MC) in the vacuum~\cite{Klevansky:1989vi,Klimenko:1990rh,Gusynin:1995nb}, and Inverse Magnetic Catalysis (IMC) around the critical temperature~\cite{Bali:2011qj,Bali:2012zg,Bali:2013esa}. Just like in a magnetic field~\cite{Miransky:2015ava}, many interesting phenomena can occur in a rotating medium as well. We could observe anomalous transport properties like chiral vortical effect~\cite{Kharzeev:2007tn,Son:2009tf,Kharzeev:2010gr}, chiral vortical wave~\cite{Jiang:2015cva} etc. These anomalous processes can be connected to experimental signals, thus can possibly be observed in experiments like HICs~\cite{Kharzeev:2015znc}. Apart from these anomalous properties, it is also interesting to observe the effect of rotation on the QCD phase diagram \textemdash\, both on the chiral and deconfinement phase transitions~\cite{Jiang:2016wvv,Chernodub:2016kxh,Wang:2018sur,Jiang:2021izj,Chernodub:2020qah,Braguta:2021jgn}. 

Here, we should keep in mind that having created a QCD medium subjected to such a tremendous amount of angular momentum gives us the liberty of exploring it with a better scope. But the existence of such rotating matter is not confined to HICs alone, rather there are many physical circumstances such as trapped non-relativistic bosonic cold atoms in condensed matter physics~\cite{Fetter:2009zz}, rapidly rotating neutron stars~\cite{Berti:2004ny} which can be a source of such rotating matter.

As compared to the magnetic field effects, the rotation-related effects are electric charge blind, and only involve kinetic properties of the QGP and strong interaction which we are mostly interested in. Experimentally, to screen out the EM effects, neutral particles with finite spin numbers are chosen as carriers of the vorticity polarisation effects. As it is difficult to detect the uncharged particles directly the distribution of their charged daughter particles serves as an alternative observable for the global polarisation effect. With the help of the $\Lambda$ polarization, the average magnitude of the vorticity of QGP had been extracted by the STAR collaboration~\cite{STAR:2017ckg}. In these measurements, the expectation of $\Lambda$ polarisation as well as the vorticity behavior of collision energy has been confirmed as well. All the results seem to be understandable by considering the energy shift induced by the vorticity polarisation to spins. However, the theory became a little vague when the $K^{*0}$ and $\phi$ mesons' measurements were presented in~\cite{ALICE:2019aid}. The mismatch between these measurements indicates the fine structure of hadrons may play a non-negligible role in polarisation processes.

From the experimental side, one important property of heavy-ion collisions is the elliptic flow $v_2$, which reflects the initial spatial anisotropy of peripheral collisions transferring into the momentum anisotropy. Another valuable electromagnetic signal such as dilepton or photon production reflects properties of the quark and gluon distributions of the QGP, the advantage of this signal is that once the dilepton and photons are produced, they will escape the medium without significant interaction. Recently it has been found from both the PHENIX experiment at RHIC \cite{PHENIX:2011oxq} and the ALICE experiment at the LHC \cite{Lohner:2012ct} that direct photons show a large elliptic flow, which can be comparable to that of hadrons. This is most puzzling because the photons carry information of early stages and the flow at early times should be small. One natural consideration is that the large photon $v_2$ or anisotropy might be induced by a large magnitude of magnetic field or rapid vorticity, which will polarize the medium and enhance the anisotropy of the system. It has been investigated how magnetic field will affect the photon elliptic flow $v_2$ in \cite{Wang:2020dsr,Wang:2021ebh}, in this manuscript, we would like to investigate the effect of rotation on dilepton rate (DR). Dilepton pairs originate from the same virtual photon\cite{Salabura:2020tou}. 

Now, the DR can be calculated from the correlation functions (CF) and their spectral representations~\cite{Forster:1975pm}. CFs and their spectral representations have been a reliable tool for exploring the many-particle systems in vacuum~\cite{Davidson:1995fq} as well as in different extreme conditions~\cite{Kapusta:2006pm,Bellac:2011kqa}. The spatial part of the spectral function (SF) can be related with the conductivity~\cite{Francis:2011bt}, whereas the temporal part can capture the response of the conserved density fluctuations~\cite{Kunihiro:1991qu}. On the other hand, the vector-vector CF and its spectral representation are associated with the dilepton production rate (DR)~\cite{Kapusta:2006pm,Bellac:2011kqa}. Investigations of the rate of such lepton pair produced in the HICs are of real importance, as having a larger mean free path than the system size the dilepton can bring along less contaminated information about the stages at which they are created.

Being a very important quantity the vector-vector current CF, its spectral representation and the DR therefrom have been extensively explored on numerous occasions in different possible scenarios. It is investigated in the ambit of effective models like Nambu\textemdash Jona-Lasinio and its Polyakov loop extended version in Ref.\cite{Islam:2014sea} and the matrix model in Refs.~\cite{Gale:2014dfa,Hidaka:2015ima}. All these calculations were performed with zero magnetic field. The SF and DR have also been calculated using a basic field theoretical approach for both imaginary~\cite{Sadooghi:2016jyf,Bandyopadhyay:2016fyd,Das:2021fma} and real time~\cite{Bandyopadhyay:2017raf,Ghosh:2018xhh} formalisms. Efforts have also been made to explore the effects of magnetic fields on SF and DR through effective model scenarios~\cite{Islam:2018sog,Ghosh:2020xwp}.

The external fields from orbital angular momentum and/or magnetic field cause spin polarization of quarks, thus induces azimuthal anisotropy naturally. Hydrodynamic simulations have studied elliptic flow of dilepton rate in different initial conditions~\cite{Xu:2014ada,Vujanovic:2013jpa,Kasmaei:2018oag,Chatterjee:2007xk}. In Au + Au collisions at $\sqrt{S_{NN}}=200$ GeV, dielectron azimuthal anisotropy at mid-rapidity was investigated by STAR collaboration \cite{STAR:2014aok}. However, it is still a lack of extensive theoretical analysis on the ellipticity of lepton pairs induced by external fields, such as strong magnetic field and vorticity, particularly.

In this paper, we mainly focus on the rotational effect on spectral function, dilepton rate and its anisotropy, which has not been investigated adequately in previous research. From the CF we find its spectral representation and thus can explore the effect of rotation on SF. A significant impact of the rotation is to increase the strength of the SF when it is varied as a function of the invariant mass scaled with temperature. Our investigation also will compare it with the impacts of other parameters like the temperature, chemical potential, etc on the SF when intertwined with the rotational effects. For a fixed value of the angular velocity, our observation is that variations of other parameters produce known effects on the SF as in the case of zero rotation.Once we have the SF, it is straightforward to obtain the DR from therein. We find that the enhancement in the SF due to the rotation effects is also reflected in the DR. We estimate the rate as a ratio to the Born rate, which is nothing but the rate for one loop photon polarization tensor in absence of rotation. This ratio, we learned, goes over unity at the low invariant mass region and can be $15-20\%$ higher than the usual Born rate. 

Furthermore, we investigate the azimuthal anisotropy of the DR induced by rotation. A non-negligible elliptic flow  $v_{2}$ is presented in this work, and its non-monotonic behavior indicates the competition between centrifugation and polarization. It is worth noticing that quark polarization introduces a negative $v_{2}$ in a large angular velocity region. Finally, we calculated the electromagnetic screening by estimating the Debye mass from the zeroth component of the polarisation tensor for a rotating QCD medium. We found that the Debye mass gets suppressed in the presence of a non-zero angular velocity. These properties are indeed important findings in terms of the dilepton signals in a non-central HIC and signify the important role that rotation could play in it. 

Our manuscript is organized as follows. In Sec.~\ref{sec:form} we outline the formalism of dilepton rate and field theory in a rotating frame. In Sec.~\ref{sec:cal}, we present the one loop polarization function under rotation and extract its explicit imaginary part. After that, the numerical result for dilepton rate, elliptic flow, and Debye mass are discussed in Sec.~\ref{sec:res}. Finally, conclusions are drawn in Sec.~\ref{sec:con}, and details of calculations are provided in appendices

\section{Formalism}
\label{sec:form}
\begin{figure}
\begin{center}
\includegraphics[scale=0.45]{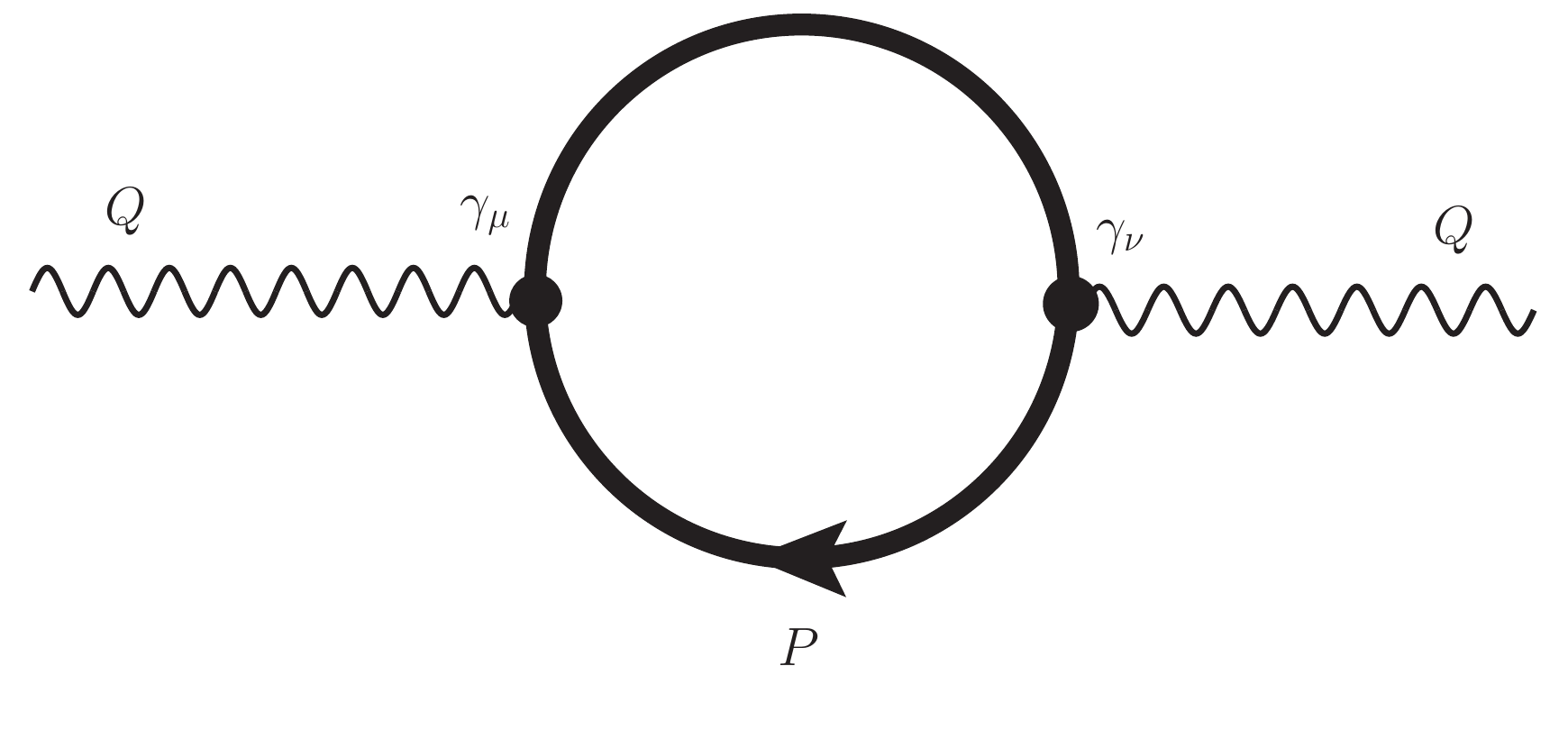} 
\caption{The photon polarization tensor.}
\label{fig:pol_tensor}
\end{center}
\end{figure}

In this present study, we have considered a QCD medium under rotation. We calculate the current-current (vector) correlation function (CF) or equivalently the photon polarisation tensor in presence of the rotation. The photon polarisation tensor is depicted in Fig.~\ref{fig:pol_tensor}. Once we have the CF, we estimate the spectral function (SF) therefrom. Then using that SF we can calculate the dilepton rate (DR) for a rotating medium. To start with the calculation, we need to write down the quark propagator in presence of rotation. Since the topic of rotating medium is relatively new, we give here a few of details about it before getting to the propagator.

To study the SF and other related phenomena in a rotating medium, we adopt a rotating frame with vierbein formalism~\cite{Yamamoto:2013zwa,Jiang:2016woz}. In a co-moving frame, vierbein(also called tetrad) can be expressed as: $e^{a}_{\ \mu}=\delta^a_{\ \mu}+\delta^a_{\ i}\delta^0_{\ \mu} \, v_i$ and $e_{a}^{\ \mu}=\delta_a^{\ \mu}-\delta_a^{\ 0}\delta_i^{\ \mu} \, v_i$ $\left(a,\mu=0,1,2,3\right.$ and $\left. i=1,2,3\right)$. For uniformly rotating frame, $\vec{v} =\vec{\Omega}\times\vec{x}$ will be adopted. As a consequence, tetrad fields give us a metric in rotating frame: $g_{\mu\nu}=\eta_{ab}e^{a}_{\ \mu}e^{b}_{\ \nu}$, where $\eta_{ab}$ is the metric of Minkowski space-time and $\gamma^{a}$ is the flat gamma matrices. Correspondingly, $\bar{\gamma}^{\mu}=e_{a}^{\ \mu}\gamma^{a}$ satisfy the Clifford algebra $\{\bar{\gamma}^{\mu},\bar{\gamma}^{\nu}\}=2 g^{\mu\nu}$ in general space-time.

The free Lagrangian for Dirac fermions in the co-rotating frame is given by~\cite{Jiang:2016woz},
\begin{equation}
\mathcal{L} = \bar{\psi}[i\bar{\gamma}^{\mu}(\partial_{\mu}+\Gamma_{\mu})-M_{f}]\psi,
\end{equation}
where $M_{f}$ is the current mass for particular flavor. The connection $\Gamma_\mu$  given by $\Gamma_\mu=\frac{1}{4}\times\frac{1}{2}[\gamma^a,\gamma ^b] \, \Gamma_{ab\mu}$ describes parallel transportation for a fermion field $\psi$, where $\Gamma_{ab\mu}=\eta_{ac}(e^c_{\ \sigma} G^\sigma_{\ \mu\nu}e_b^{\ \nu}-e_b^{\ \nu}\partial_\mu e^c_{\ \nu})$ is called spin connection. Spin connection is determined by the tetrad field uniquely, because $G^\sigma_{\ \mu\nu}$ is the Christoffel connection associated with metric $g_{\mu\nu}=\eta_{ab}e^{a}_{\ \mu}e^{b}_{\ \nu}$~\cite{Yamamoto:2013zwa,Jiang:2016wvv}. 

\begin{figure}
\begin{center}
\includegraphics[scale=0.35]{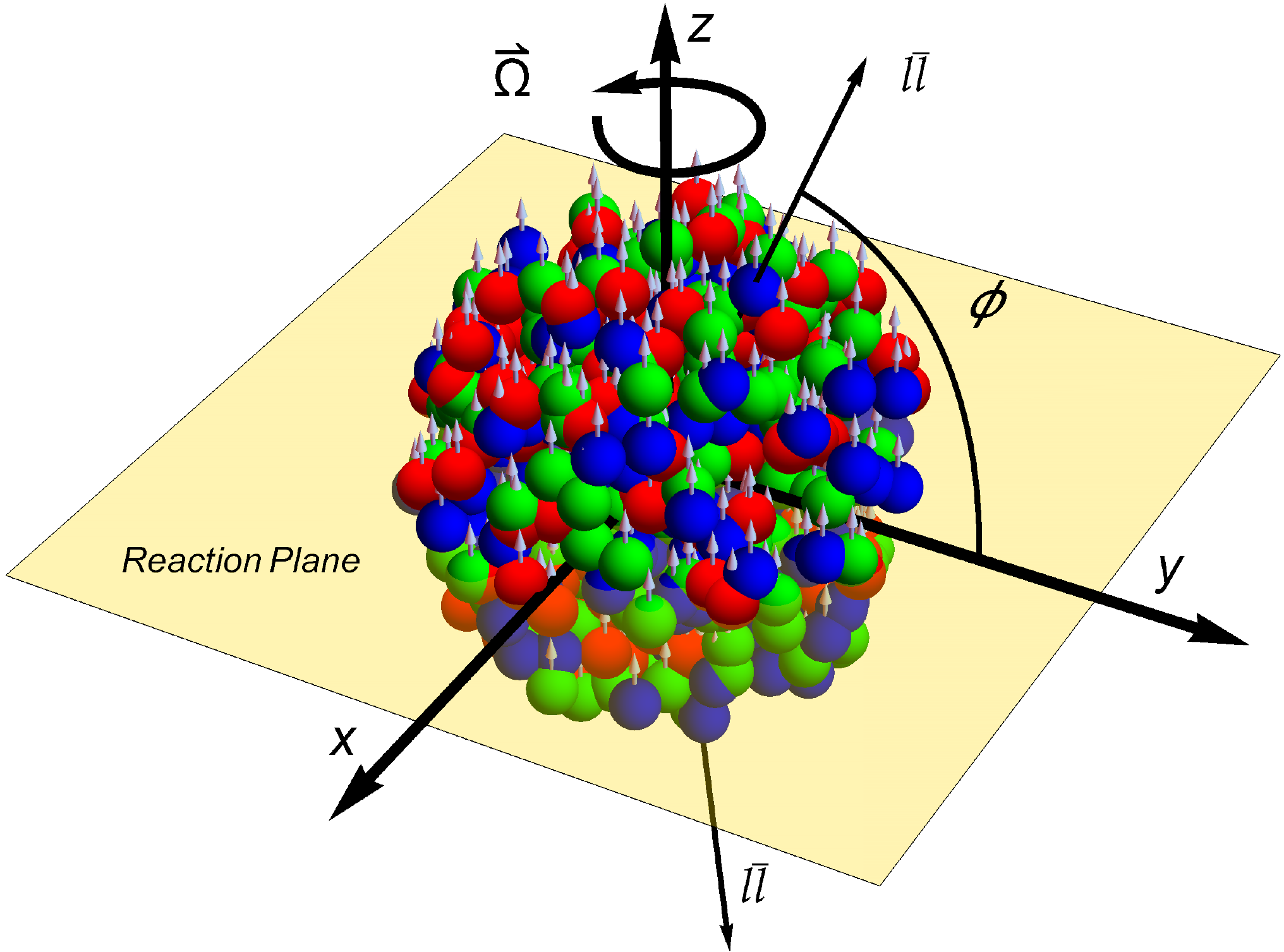} 
\caption{An artist's impression for the dilepton production from a rotating QGP.}
\label{fig:demonstation}
\end{center}
\end{figure}

In heavy-ion collisions as demonstrated in Fig.~(\ref{fig:demonstation}), it is easy to check that the direction of rotation or the total angular momentum is along the out-of-plane direction, i.e., perpendicular to the reaction plane~\cite{Jiang:2016woz,Deng:2016gyh}. In this work, reaction plane is represented as $x$-$y$ plane where $x$-axis goes along the beam direction and $y$-axis goes along the impact parameter. Consequently, the direction of rotation is along the $z$-axis. People can calculate the nonzero terms of spin connection $\Gamma_{ab\mu}$ and combine it with $\bar{\gamma}^{\mu}=e_{a}^{\ \mu}\gamma^{a}$. Finally, the free Lagrangian with finite chemical potential under rotation is given by~\cite{Wang:2018sur}
\begin{equation}
\mathcal{L}=\bar{\psi}[i\gamma^{a}\partial_{a}+\gamma^{0}(\Omega \hat{J_{z}}+\mu)-M_{f}]\psi,
\end{equation}
where $\mu$ is chemical potential and $J_{z}$ is the third component of the total angular momentum $\vec{J}=\vec{x}\times\vec{p}+\vec{S}$. Here, we can extract the third component from the spin operator $\vec{S}=\frac{1}{2}\left(
                      \begin{array}{cc}
                        \vec{\sigma} & 0 \\
                        0 & \vec{\sigma} \\
                      \end{array}
                    \right)
$. It is seen that the angular velocity plays a similar role as the chemical potential.

With all these details, we can now write down the quark propagator in presence of rotation as~\cite{Wei:2020xfd},
\begin{equation}
\label{eq:quark_prop}
\begin{aligned}
S(\tilde{r};\tilde{r}')&=\frac{1}{(2 \pi )^2}\sum _n\int\frac{d k_0}{2\pi } \int k_t d k_t \int d k_z \frac{e^{i n \left(\theta -\theta '\right)}e^{-i k_0 \left(t-t'\right)+i k_z \left(z-z'\right)}}{[k_{0}+(n+\frac{1}{2})\Omega]^{2}-k_{t}^{2}-k_{z}^{2}-M_{f}^{2}+i\epsilon} \\
&
\times\left\{\left[[k_{0}+(n+\frac{1}{2})\Omega]\gamma^{0}-k_{z}\gamma^{3}+M_{f}\right]\left[J_{n}(k_{t}r)J_{n}(k_{t}r')\mathcal{P}_{+}+e^{i (\theta-\theta')}J_{n+1}(k_{t}r)J_{n+1}(k_{t}r')\mathcal{P}_{-}\right]\right.\\
&-i\left.\gamma^{1}k_{t}e^{i \theta}J_{n+1}(k_{t}r)J_{n}(k_{t}r')\mathcal{P}_{+}-\gamma^{2}k_{t}e^{-i\theta '}J_{n}(k_{t}r)J_{n+1}(k_{t}r')\mathcal{P}_{-}
\right\},
\end{aligned}
\end{equation}
where $\mathcal{P}_{\pm}=\frac{1}{2}(1\pm i\gamma^{1}\gamma^{2})$ are projection operators and $\tilde{r}=(t,r,\theta,z)$ is a point in spacetime. Here we will use $\Omega$ to represent angular velocity to distinguish it from real energy $\omega$. In current discussion, we consider $r\ll \Omega^{-1}$ so that causality will be  ensured. In real heavy-ion physical process, $k_t$ will be in a finite range. So, $k_{t}r\ll 1$ will cause that the contribution of higher order Bessel functions are suppressed severely. In fact, z-angular momentum quantum number $n=0,\pm 1$ will contribute almost entire rotation effect in this framework. 

Now, using this general fermionic propagator in a rotating medium in Eq.~\eqref{eq:quark_prop}, the vector current-current CF~\cite{Islam:2014sea}, relevant to the one loop photon self-energy diagram in Fig.~(\ref{fig:pol_tensor}), can be expressed as:
\begin{equation}
\Pi^{ab}(q)=-i \int d^{4}\tilde{r}Tr_{Dfc}[i \gamma^{a}S(0;\tilde{r})i \gamma^{b}S(\tilde{r};0)]e^{i q\cdot \tilde{r}},
\label{eq:cf}
\end{equation}
where $q$ is the 4-momentum of external photon line (Fig.~\ref{fig:pol_tensor}) and the trace is over Dirac (D), flavour (f) and colour (c) spaces. The traces over the flavour and colour spaces can be performed and the above expression~\eqref{eq:cf} can be rewritten as,
\begin{equation}
\Pi^{ab}(q)=-i N_f N_c \int d^{4}\tilde{r}Tr_{D}[i \gamma^{a}S(0;\tilde{r})i \gamma^{b}S(\tilde{r};0)]e^{i q\cdot \tilde{r}},
\label{eq:cf_tr}
\end{equation}
where $N_f$ and $N_c$ are the numbers of flavour and colour, respectively; the only trace remaining is the one in the Dirac space. It is to be noted that the CF in Eq.~\eqref{eq:cf_tr} can be easily related to the photon polarisation tensor by appropriately inserting the coupling constant. The SF is also related to the imaginary part of the CF, as
\begin{equation}
\sigma_V(q)=\frac{1}{\pi}{\mathrm{Im}}\Pi^a_a(q),
\label{eq:sf}
\end{equation}
where the Lorentz index $a$ is contracted over. This SF is a quantity of great importance. Many key observables like DR, electrical conductivity, Debye mass etc can be extracted from it. In this work, for a rotating QCD medium, after estimating the SF and discussing it in detail, we extract the DR from it. 

The DR can be related to the SF through the well-known relation~\cite{Islam:2014sea},
\begin{align}
\frac{dN}{d^4xd^4q} = \frac{5\alpha^2}{27\pi^2} 
\frac{1}{M^2}\frac{1}{{\mathrm{ex}p}(\frac{\omega}{T})-1}\sigma_V(q),
\label{eq:dr}
\end{align}
where the invariant mass of the lepton pair $M^2=A^2=\omega^2-q^2$. Eqs.~\eqref{eq:sf} and~\eqref{eq:dr} are two major equations that we evaluate in the next subsection and discuss in detail in the result section.

\section{Calculation}
\label{sec:cal}
\subsection{Diagonal components of the CF}
\label{ssec:com_cf}
In this subsection, we start with calculating the components of the CF, which will eventually give us the total of it. We insert the propagator~\eqref{eq:quark_prop} into the CF Eq.~\eqref{eq:cf_tr} and then simplify to obtain the diagonal components. The $00$ component of the CF is given as,
\begin{equation}
\begin{aligned}
\Pi^{00}(q)&=
-2N_{f}N_{c}\sum_{\eta=\pm 1}\int \frac{d^{4} p}{(2\pi)^{4}} 
\frac{M_{f}^2+\left(p_{0}+\frac{\eta\Omega}{2}\right) \left(p_{0}+q_{0}+\frac{\eta\Omega}{2}\right)+ \vec{p}\cdot(\vec{p}+\vec{q})}
{\left[\left(p_{0}+\frac{\eta\Omega }{2}\right)^2-\vec{p}^2-M_{f}^2\right] \left[\left(p_{0}+q_{0}+\frac{\eta\Omega }{2}\right)^2-(\vec{p}+\vec{q})^2-M_{f}^2\right]},
\end{aligned}
\end{equation}
where we will put $N_{f}=2$ and $N_{c}=3$. Similarly, we can obtain the spatial components as well. Then, the finite temperature version of polarisation function can be obtained by replacing $p_{0}$ with Matsubara frequency. The explicit expressions of those components are  cumbersome, so they will be included in Appendix.\ref{appendix:a}.

After Matsubara Summation calculation with angular velocity $\Omega$ in Appendix.\ref{appendix:b}, we obtain simplified expressions for polarisation functions. The $00$ component can be read as:  
\begin{equation}
\begin{aligned}
\Pi^{00}(\omega,\vec{q})&=
\frac{1}{2} N_{f}N_{c}\sum_{\eta=\pm 1}\int \frac{d^{3}\vec{p}}{(2\pi)^{3}} \\
&\times\frac{1}{E_{p}E_{k}}\left\{\frac{E_{p}E_{k}+\vec{p}\cdot\vec{k}+M_{f}^{2}}{\omega-E_{p}+E_{k}}\right.\\
&\left[f(E_{k}-\mu-\frac{\eta\Omega}{2})-f(E_{p}-\mu+\frac{\eta\Omega}{2})-f(E_{p}+\mu+\frac{\eta\Omega}{2})+f(E_{k}+\mu-\frac{\eta\Omega}{2})\right]\\
&+\left[E_{p}E_{k}-\vec{p}\cdot\vec{k}-M_{f}^{2}\right]\left(\frac{1}{\omega-E_{p}-E_{k}}-\frac{1}{\omega+E_{p}+E_{k}}\right)\\
&\times\left.\left[1-f(E_{p}-\mu-\frac{\eta\Omega}{2})-f(E_{k}+\mu+\frac{\eta\Omega}{2})\right]\right\},
\label{eq:pi_00}
\end{aligned}
\end{equation}
where $k=p+q$ and $\omega$ is analytically extended real energy. Comparing it with the ordinary form for 00-component in Ref.(\cite{Islam:2014sea}), we observe that the rotation effect is embedded in the distribution functions, and the angular velocity $\Omega$ acts as an effective chemical potential. However, the rotation effect will affect the spatial part profoundly. It will affect the transverse part and longitudinal part in different ways. Now, the transverse part can be read as:  
\begin{equation}
\label{eq:transversepart}
\begin{aligned}
\Pi^{11}(\omega,\vec{q})+\Pi^{22}(\omega,\vec{q})&=
\frac{1}{2} N_{f}N_{c}\sum_{\eta=\pm 1}\int \frac{d^{3}\vec{p}}{(2\pi)^{3}} \\
&\times\frac{1}{E_{p}E_{k}}\left\{\frac{2E_{p}E_{k}-2 p_{z}(p_{z}+q_{z})-2M_{f}^{2}}{\omega-E_{p}+E_{k}-\eta\Omega}\right.\\
&\times\left[f(E_{k}-\mu-\frac{\eta\Omega}{2})-f(E_{p}-\mu+\frac{\eta\Omega}{2})-f(E_{p}+\mu+\frac{\eta\Omega}{2})+f(E_{k}+\mu-\frac{\eta\Omega}{2})\right]\\
&+\left[2E_{p}E_{k}+2 p_{z}(p_{z}+q_{z})+2M_{f}^{2}\right]\left(\frac{1}{\omega-E_{p}-E_{k}-\eta\Omega}-\frac{1}{\omega+E_{p}+E_{k}+\eta\Omega}\right)\\
&\times\left.[1-f(E_{p}-\mu+\frac{\eta\Omega}{2})-f(E_{k}+\mu+\frac{\eta\Omega}{2})]\right\}.
\end{aligned}
\end{equation}
Similarly, the longitudinal part can be read as: 
\begin{equation}
\begin{aligned}
\Pi^{33}(\omega,\vec{q})&=
\frac{1}{2} N_{f}N_{c}\sum_{\eta=\pm 1}\int \frac{d^{3}\vec{p}}{(2\pi)^{3}} \\
&\times\frac{1}{E_{p}E_{k}}\left\{\frac{E_{p}E_{k}-[+p_{x}(p_{x}+q_{x})+ p_{y}(p_{y}+q_{y})- p_{z}(p_{z}+q_{z})]-M_{f}^{2}}{\omega-E_{p}+E_{k}}\right.\\
&\times\left[f(E_{k}-\mu-\frac{\eta\Omega}{2})-f(E_{p}-\mu-\frac{\eta\Omega}{2})-f(E_{p}+\mu+\frac{\eta\Omega}{2})+f(E_{k}+\mu+\frac{\eta\Omega}{2})\right]\\
&+\left[E_{p}E_{k}+[p_{x}(p_{x}+q_{x})+ p_{y}(p_{y}+q_{y})- p_{z}(p_{z}+q_{z})]+M_{f}^{2}\right]\left(\frac{1}{\omega-E_{p}-E_{k}}-\frac{1}{\omega+E_{p}+E_{k}}\right)\\
&\times\left.\left[1-f(E_{p}-\mu-\frac{\eta\Omega}{2})-f(E_{k}+\mu+\frac{\eta\Omega}{2})\right]\right\}.
\end{aligned}
\end{equation}
Firstly, we should compare transverse and longitudinal parts. We can find an additional $\Omega$ term in the denominator of Eq.~\eqref{eq:transversepart}, so it is infeasible to merge the transverse part and longitudinal part as usual. Furthermore, the difference will induce anisotropy of dilepton production in the transverse plane.

\subsection{Imaginary parts}
\label{ssec:com_cf_im}
After further simplifications, the details of which can be found in the appendix, we can write down the temporal component of the imaginary part of the self energy. The $00$ part can be written as
\begin{equation}
 \begin{aligned}
Im\Pi^{00}(\omega,\vec{q})&=
\frac{1}{8\pi} N_{f}N_{c}\sum_{\eta=\pm 1}\int_{p_{-}}^{p_{+}}p d p\frac{4\omega E_{p}-4E_{p}^{2}-M^{2}}{2q E_{p}}[f(E_{p}-\mu-\frac{\eta\Omega}{2})+f(E_{p}+\mu-\frac{\eta\Omega}{2})-1].
\end{aligned}
\end{equation}
Here, $p_{\pm}=\pm\frac{q}{2}+\frac{\omega}{2}\sqrt{1-\frac{4M_{f}^2}{\omega^2-q^{2}}}$ constrain the integral region. From now on, $q$ also represents $q=|\vec{q}|$ for convenience. Angular velocity $\Omega$ merely appears in the distribution functions, which indicates its impact as an effective chemical potential.

The other three components are calculated in Appendix.\ref{appendix:c}, which are given as:
\begin{equation}
\begin{aligned}
&Im[\Pi^{11}(\omega,\vec{q})+\Pi^{22}(\omega,\vec{q})]\\
&=-\frac{1}{2}\pi N_{f}N_{c}\sum_{\eta=\pm 1}\int_{p^{\Omega}_{-}}^{p^{\Omega}_{+}} \frac{p d p}{(2\pi)^{2}}\\
&\times\left\{+\frac{2E_{p}(\omega-E_{p}+\eta\Omega)+[(3\frac{q_{z}^2}{q^{2}}-1)(p\cos{\theta_{1}})^{2}
+p^{2}(1-\frac{q_{z}^{2}}{q^{2}})+2 \frac{q_{z}^{2}}{q}p\cos{\theta_{1}}]+2 M_{f}^{2}}{q E_{p}}\right.
\\
&\times\left.[1-f(E_{p}-\mu-\frac{\eta\Omega}{2})-f(E_{p}+\mu-\frac{\eta\Omega}{2})]\right\}.
\end{aligned}
\end{equation}
Here, $p^{\Omega}_{\pm}=\pm\frac{q}{2}+\frac{\omega+\eta\Omega}{2}\sqrt{1-\frac{4M_{f}^2}{(\omega+\eta\Omega)^2-q^{2}}}$ constrain the integral region and $\cos{\theta_{1}}=\frac{(\omega+\eta\Omega)^{2}-2(\omega+\eta\Omega)\sqrt{p^{2}+M_{f}^{2}}-q^{2}}{2pq}$ stands for angle between $\vec{p}$ and $\vec{q}$. We can find that the rotation affects $p^{\Omega}_{\pm}$ and $\cos{\theta_{1}}$ terms. 

Similarly, the longitudinal component can be read as: 
\begin{equation}
\begin{aligned}
&Im\Pi^{33}(\omega,\vec{q})\\
&=-\frac{1}{2}\pi N_{f}N_{c}\sum_{\eta=\pm 1}\int_{p_{-}}^{p_{+}} \frac{p d p}{(2\pi)^{2}}\\
&\times\left\{
\frac{E_{p}(\omega-E_{p})+ \frac{1}{2}[(3\frac{q_{x}^{2}+q_{y}^{2}-q_{z}^2}{q^{2}}-1)(p\cos{\theta_{0}})^{2}+p^{2}(1-\frac{q_{x}^{2}+q_{y}^{2}-q_{z}^{2}}{q^{2}})+2 \frac{q_{x}^{2}+q_{y}^{2}-q_{z}^{2}}{q}p\cos{\theta_{0}}]+M_{f}^{2}}{q E_{p}}\right.
\\
&\times\left.[1-f(E_{p}-\mu-\frac{\eta\Omega}{2})-f(E_{p}+\mu-\frac{\eta\Omega}{2})]\right\}.
\end{aligned}
\end{equation}
Here, $p_{\pm}=\pm\frac{q}{2}+\frac{\omega}{2}\sqrt{1-\frac{4M_{f}^2}{\omega^2-q^{2}}}$ constrain the integral region and $\cos{\theta_{0}}=\frac{\omega^{2}-2\omega\sqrt{p^{2}+M_{f}^{2}}-q^{2}}{2pq}$.
It's obvious and profound that $Im\Pi^{a}_{a}(\omega,\vec{q})$ can be expressed as $Im\Pi^{a}_{a}(\omega,|\vec{q}|,q_{z})$. Here, rotation effect has broken the spatial symmetry, and spectral function and corresponding dilepton production rate will rely on both the magnitude and azimuthal angle of momentum $\vec{q}$. 

\subsection{Ellipticity of dilepton}
\label{ssec:com_dilepton_v2}
As we mentioned in previous sections, rotation effect will induce the anisotropy of spectral function and dilepton production rate. In this section, we will investigate the ellipticity of dilepton in transverse plane(y-z plane). Elliptic flow coefficients can be defined as~\cite{Wang:2020dsr,Voloshin:1994mz}:
\begin{equation}
E \frac{\mathrm{d}^{3} N}{\mathrm{~d}^{3} \mathbf{q}}=\frac{1}{2 \pi} \frac{\mathrm{d}^{2} N}{q_{T} \mathrm{~d} q_{T} \mathrm{~d} y}\left(1+2 \sum_{n=1}^{\infty} v_{n} \cos \left[n\left(\phi-\Psi_{\mathrm{RP}}\right)\right]\right),
\end{equation}
where, $q_{T}=\sqrt{q_{y}^{2}+q_{z}^{2}}$ is the transverse momentum and $\Psi_{RP}$ is the reaction plane angle. Capital T indicates that quantities are perpendicular to the beam direction. The beam direction is chosen as the x-axis, which is also considered as the polar axis and $\phi$ is the azimuthal angle. Harmonic coefficients $v_{n}$ can be calculated by~\cite{Xu:2014ada}:
\begin{equation}
v_{n}\left(q_{T}\right)=\frac{\int_{0}^{2 \pi} d \phi \frac{d N}{q_{T} d q_{T} d \phi d y} \cos \left(n\phi\right)}{\int_{0}^{2 \pi} d \phi \frac{d N}{q_{T} d q_{T} d \phi d y}}.
\end{equation}
Here, we overlook the difference of reaction plane and event plane. Elliptic flow coefficient $v_{2}$ is calculated in mid-rapidity, i.e. $q_{x}=0$ and $\frac{q_{z}}{q_{T}}=\sin \phi$. We can divide $Im\Pi_{ii}$ into $\phi$ dependent plus independent parts.
$Im\Pi^{00}(\omega,\vec{q})$ is totally $\phi$ independent, while $Im[\Pi^{11}(\omega,\vec{q})+\Pi^{22}(\omega,\vec{q})]$ can be divided into:
\begin{equation}
\begin{aligned}
(Im[\Pi^{11}(\omega,\vec{q})+\Pi^{22}(\omega,\vec{q})])_{dep}
&=-\frac{1}{2}\pi N_{f}N_{c}\cos 2\phi\sum_{\eta=\pm 1}\int_{p^{\Omega}_{-}}^{p^{\Omega}_{+}} \frac{p d p}{(2\pi)^{2}}\\
&\times\left\{\frac{-\frac{3}{2}(p\cos{\theta_{1}})^{2}
+\frac{p^{2}}{2}-pq\cos{\theta_{1}}}{q E_{p}}\right.
\times\left.[1-f(E_{p}-\mu-\frac{\eta\Omega}{2})-f(E_{p}+\mu-\frac{\eta\Omega}{2})]\right\}
\end{aligned}
\end{equation}
and
\begin{equation}
\begin{aligned}
(Im[\Pi^{11}(\omega,\vec{q})+\Pi^{22}(\omega,\vec{q})])_{ind}&=-\frac{1}{2}\pi N_{f}N_{c}\sum_{\eta=\pm 1}\int_{p^{\Omega}_{-}}^{p^{\Omega}_{+}} \frac{p d p}{(2\pi)^{2}}\\
&\times\left\{\frac{2E_{p}(\omega-E_{p}+\eta\Omega)+[\frac{1}{2}(p\cos{\theta_{1}})^{2}
+\frac{1}{2}p^{2}+pq\cos{\theta_{1}}]+2 M_{f}^{2}}{q E_{p}}\right.
\\
&\times\left.[1-f(E_{p}-\mu-\frac{\eta\Omega}{2})-f(E_{p}+\mu-\frac{\eta\Omega}{2})]\right\}.
\end{aligned}
\end{equation}
The first equation represents the $\phi$ dependent part and the second one represents the independent part. In the same manner, we can divide $Im\Pi_{33}$ into $\phi$ dependent and independent parts:
\begin{equation}
\begin{aligned}
Im\Pi^{33}(\omega,\vec{q})_{dep}
&=-\frac{1}{2}\pi N_{f}N_{c}\cos2\phi\sum_{\eta=\pm 1}\int_{p_{-}}^{p_{+}} \frac{p d p}{(2\pi)^{2}}
\left\{
\frac{[\frac{3}{2}(p\cos{\theta_{0}})^{2}-\frac{1}{2}p^{2}+pq\cos{\theta_{0}}]}{q E_{p}}\right.
\\
&\times\left.[1-f(E_{p}-\mu-\frac{\eta\Omega}{2})-f(E_{p}+\mu-\frac{\eta\Omega}{2})]\right\}.
\end{aligned}
\end{equation}
\begin{equation}
\begin{aligned}
Im\Pi^{33}(\omega,\vec{q})_{ind}
&=-\frac{1}{2}\pi N_{f}N_{c}\sum_{\eta=\pm 1}\int_{p_{-}}^{p_{+}} \frac{p d p}{(2\pi)^{2}}
\left\{
\frac{E_{p}(\omega-E_{p})+ [-\frac{1}{2}(p\cos{\theta_{0}})^{2}+\frac{1}{2}p^{2}]+M_{f}^{2}}{q E_{p}}\right.
\\
&\times\left.[1-f(E_{p}-\mu-\frac{\eta\Omega}{2})-f(E_{p}+\mu-\frac{\eta\Omega}{2})]\right\}.
\end{aligned}
\end{equation}
Combined with the definition of elliptic flow coefficient, one can easily check that:
\eq
v_{2}=\frac{\int_{0}^{2\pi}\cos2\phi d \phi [-(Im[\Pi^{11}(\omega,\vec{q})+\Pi^{22}(\omega,\vec{q})])_{\phi}-Im\Pi^{33}(\omega,\vec{q})_{\phi}]}
{\int_{0}^{2\pi} d\phi [Im\Pi^{00}(\omega,\vec{q})-(Im[\Pi^{11}(\omega,\vec{q})+\Pi^{22}(\omega,\vec{q})])_{in}-Im\Pi^{33}(\omega,\vec{q})_{in}]}.
\eeq
Actually, $\phi$-dependent part is much smaller than the independent one, so we can focus on the denominator. However, we will evaluate the whole expression of $v_{2}$ while obtaining the numerical result.
\section{Results}
\label{sec:res}
\subsection{Spectral function}
\label{ssec:res_sf}
\begin{figure}
\begin{center}
 \includegraphics[scale=0.5]{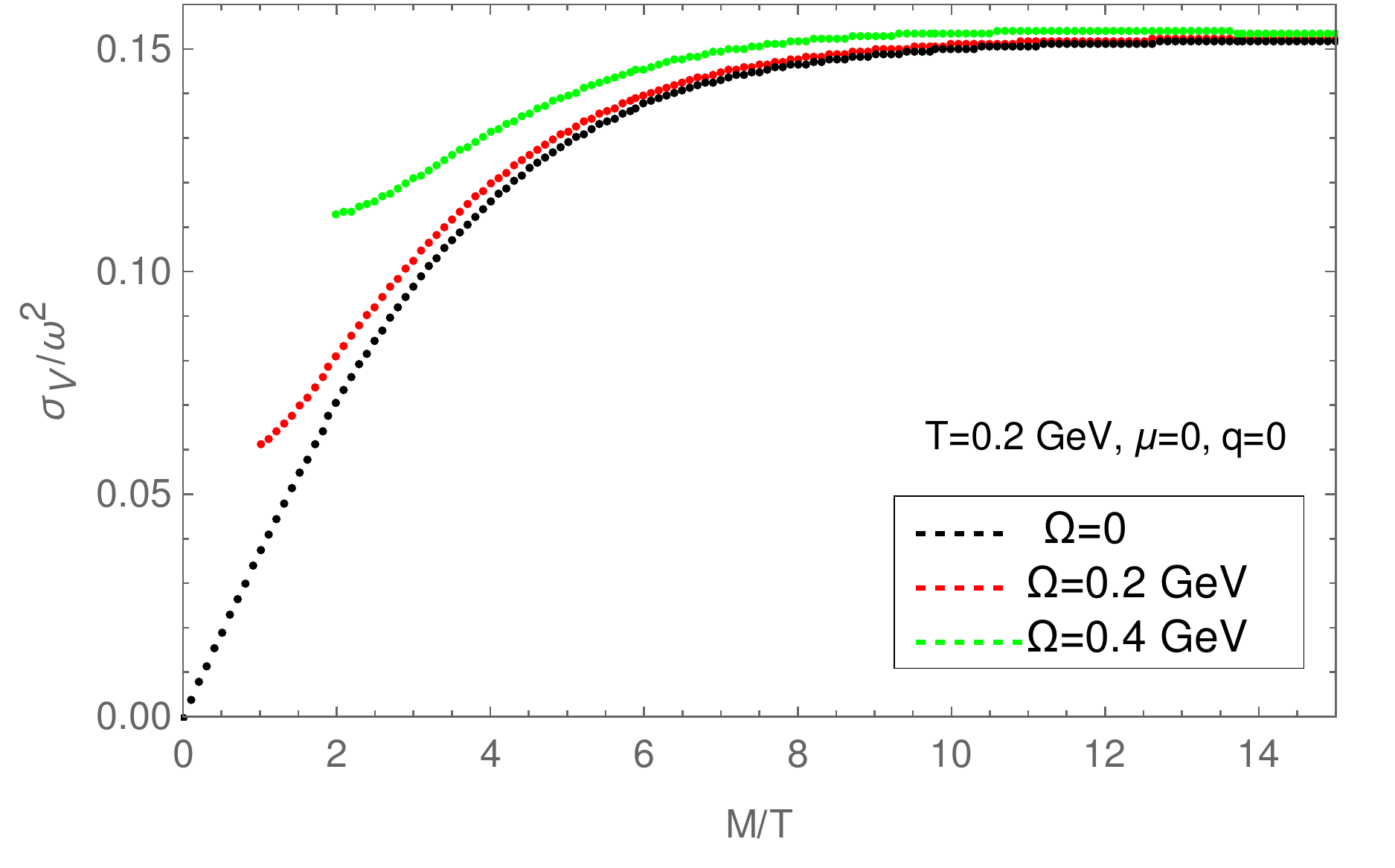}
\caption{Spectral function as a function of temperature scaled invariant mass for different values of rotation. The temperature, chemical potential and the external momentum are fixed.}
\label{fig:sf_com_diffOmega}
\end{center}
\end{figure}
In this subsection, we will discuss the spectral function (SF) in presence of rotation while taking different combinations of other parameter values. SF is an important quantity by calculating which we can shed light on other important physical quantities.

In Fig.~\ref{fig:sf_com_diffOmega} we have displayed the SF as a function of temperature scaled invariant mass $(M/T)$ for different values of the angular velocity, whereas the other parameters are kept fixed. We have used three different values of $\Omega$ $(0,\,0.2\,{\mathrm{and}}\,0.4\;{\mathrm{GeV}})$ with temperature and chemical potential kept fixed at $0.2$ GeV and $0$, respectively. This plot is obtained for zero external momentum. The black dotted line is for zero rotation and as the rotation is introduced the strength of the SF is increased, particularly at lower values of the invariant mass. The higher the values of the rotation (i.e., the angular velocity) the greater the strength of the SF. At high enough values of the invariant mass, the SFs for all values of $\Omega$ merge. The increase of the strength in the SF for non-zero values of angular velocities will have an impact on the calculation of the dilepton rate, which is discussed in the next subsection.
\begin{figure}
\begin{center}
 \includegraphics[scale=0.45]{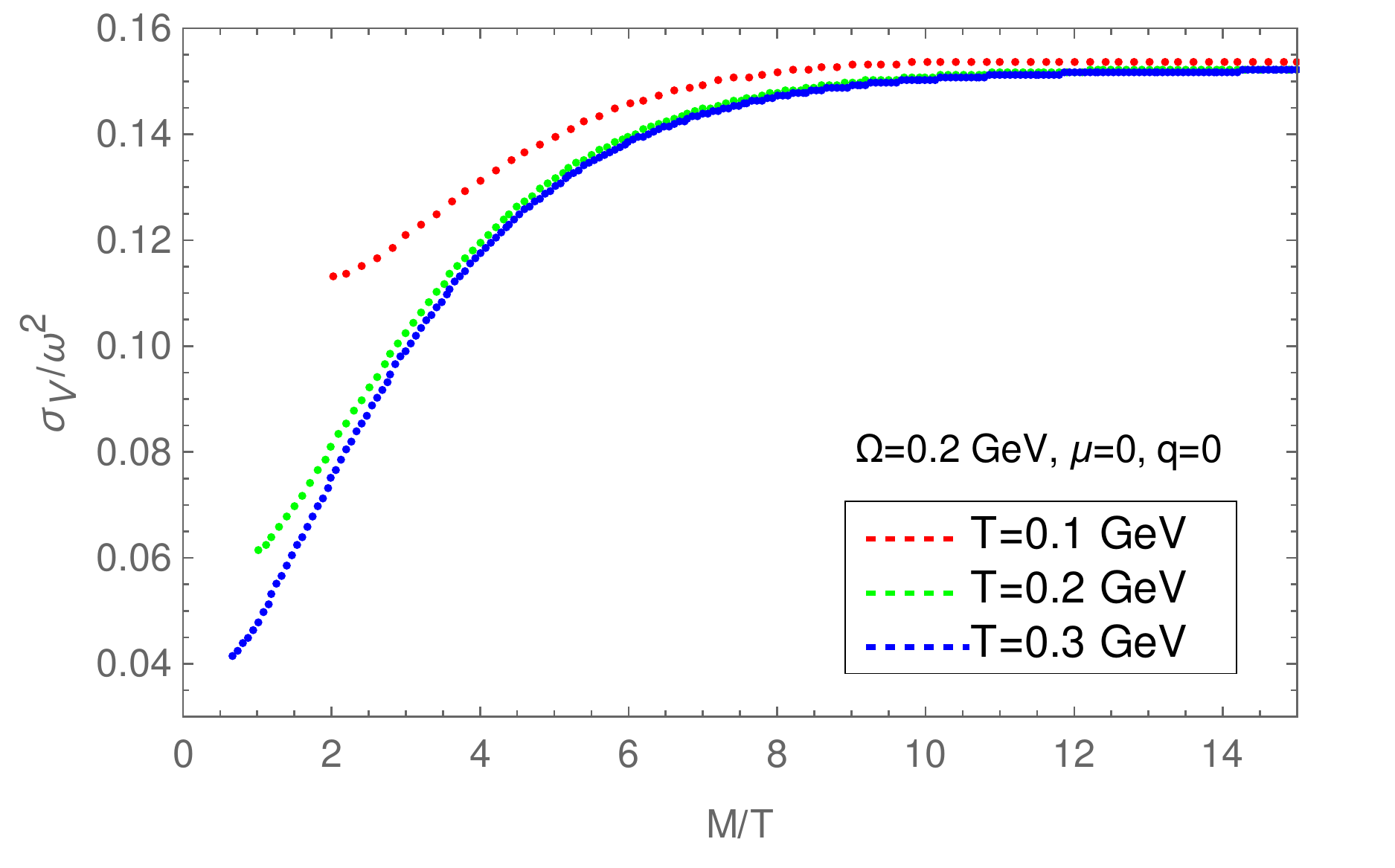}
 \includegraphics[scale=0.45]{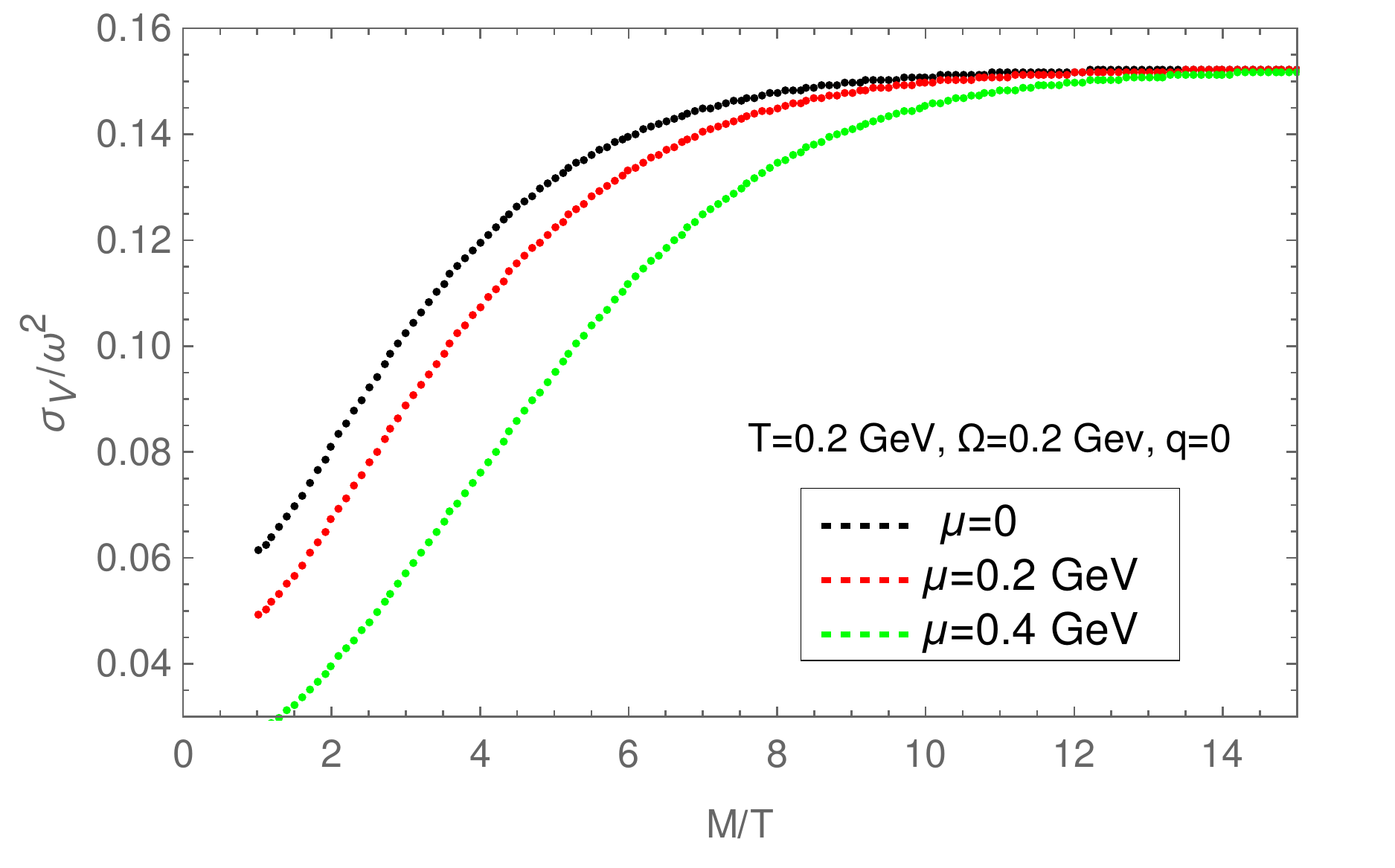}
\caption{Left panel: SF as a function of temperature scaled invariant mass for different values of temperature and for fixed values of the angular velocity, chemical potential and the external momentum. Right panel: SF for different values of chemical potential with $T$, $\Omega$ and $q$ being fixed.}
\label{fig:sf_com_diffTMu}
\end{center}
\end{figure}

In Fig.~\ref{fig:sf_com_diffTMu}, we have varied the external parameters like temperature and chemical potential, whereas angular velocity is kept fixed at a non-zero value of $0.2$ GeV. In the left panel, we have obtained the SF for different values of $T$ and for different values of $\mu$ in the right panel. For both the plots, we kept the external momentum to be zero.

This figure (\ref{fig:sf_com_diffTMu}) is, particularly, drawn with the aim of checking our results for SF with the variations of $T$ and $\mu$ for given values of $\Omega$. Here we observe, in the left panel, that as we increase the temperature for given values of other parameters, the SF decreases as a function of $M/T$. For high enough values of $M/T$, the SFs with different values of $T$ merge. This behaviour is similar to the zero rotation case. In the same way, we notice that (in the right panel) increasing the values of chemical potential decreases the strength of the SF for other given parameters. This result also conforms with the result for zero rotation where the strength of the SF always decreases with the increase of $\mu$.
\begin{figure}
\begin{center}
 \includegraphics[scale=0.5]{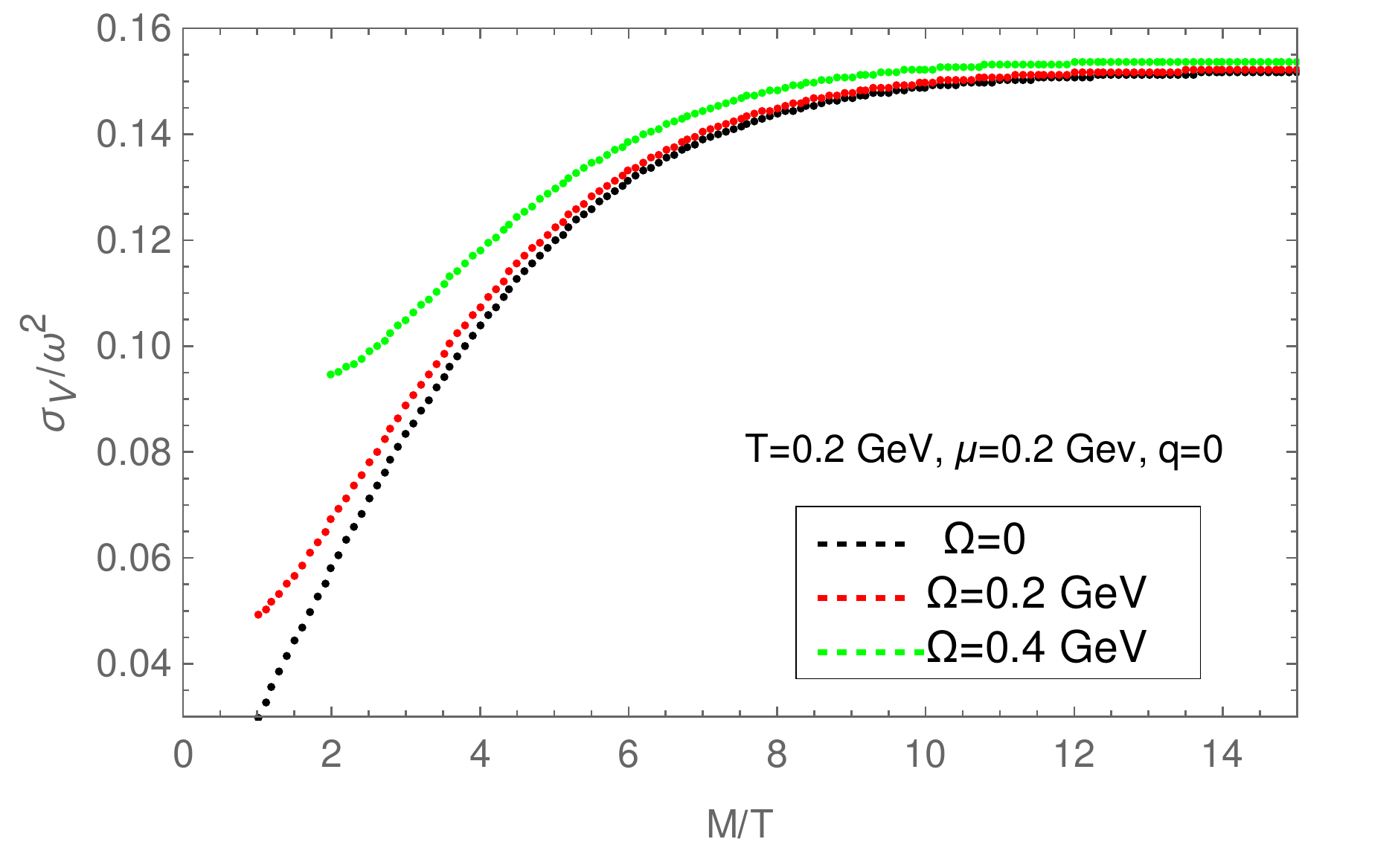}
 \caption{Spectral function as a function of temperature scaled invariant mass for different values of angular velocity, with nonzero $T$ and $\mu$.}
 \label{fig:sf_com_diffOmega_nzMu}
\end{center} 
\end{figure}

Finally, in the discussion of SF, we draw the plot for both non-zero values of $\Omega$ and $\mu$ shown in Fig.~\ref{fig:sf_com_diffOmega_nzMu}, particularly as they have contrasting effects on SF as displayed in Fig.~\ref{fig:sf_com_diffOmega} and the right panel of Fig.~\ref{fig:sf_com_diffTMu}. For this plot, we have used both $T$ and $\mu$ $=0.2$ GeV and kept the external momentum at zero. It looks qualitatively similar to Fig.~\ref{fig:sf_com_diffOmega}, that is as we increase the strength of the angular velocity the SF increases for smaller values of invariant mass and merge with the zero rotation SF at sufficiently high values.
 
\subsection{Dilepton rate}
\label{ssec:res_dr}
\begin{figure}
\includegraphics[scale=0.5]{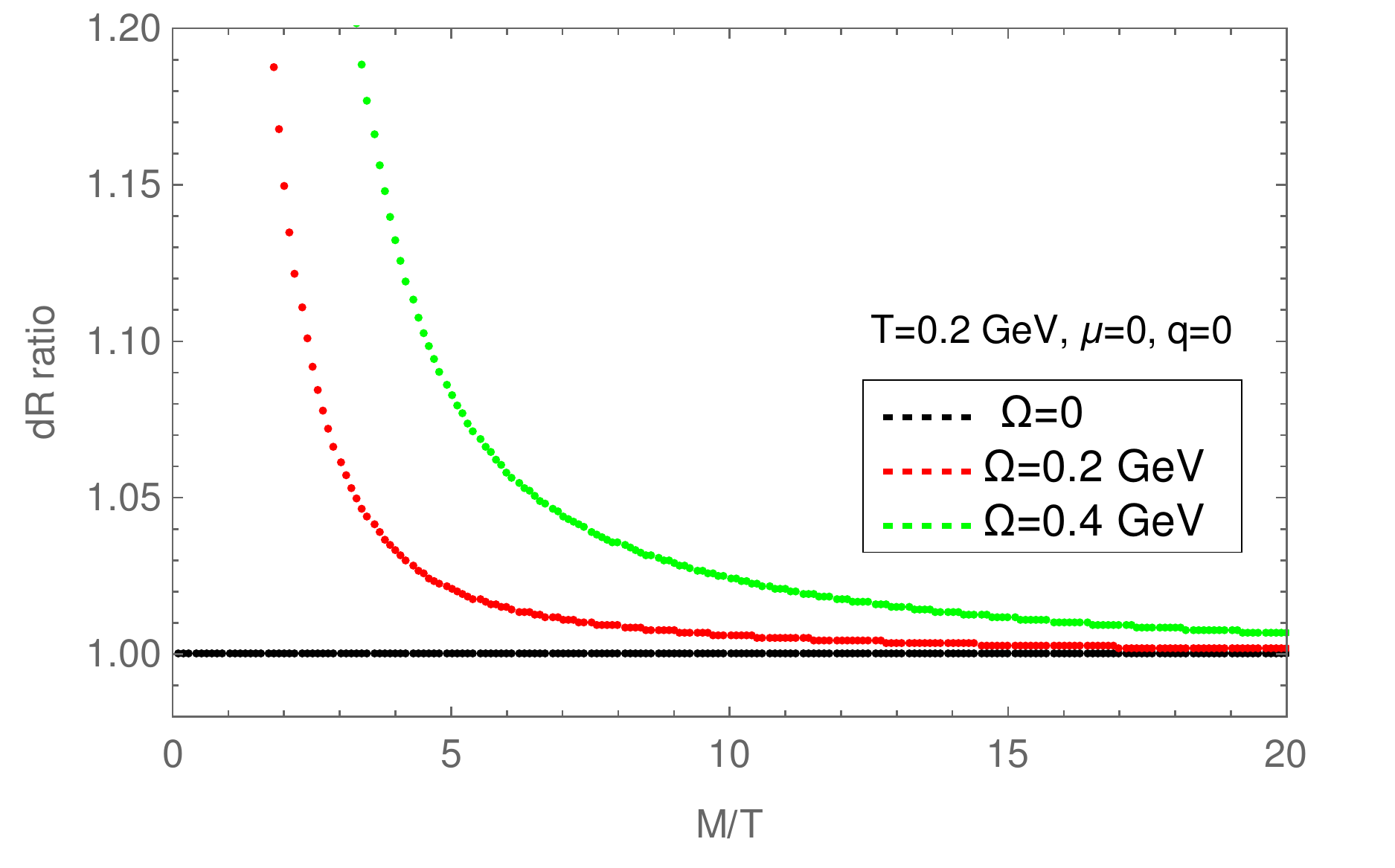}
\caption{Ratios of the dilepton rate as a function of temperature scaled invariant mass for different values of rotation. The temperature, chemical potential and the external momentum are fixed.}
\label{fig:dR_ratio_diffOmega}
\end{figure}
In this subsection, we describe the results for dilepton rate (DR) which is obtained using the SF obtained in the previous subsection. We are mainly interested here to explore the effect of rotation on the DR. Along with that we also investigate the effects of other parameters on it.

In Fig.~\ref{fig:dR_ratio_diffOmega} we have given the plot for DR as a function of $M/T$ for different values of angular velocity.  We have used three different values of $\Omega$ as $0$, $0.2$ and $0.4$ GeV, which are represented by the black, red and green points, respectively. Other parameters like temperature, chemical potential and external momentum are kept at fixed values as mentioned in the figure itself. Note that this figure is equivalent to the SF plot in Fig.~\ref{fig:sf_com_diffOmega} in terms of the choices of parameters.

The DR is plotted as a ratio between DR at non-zero $\omega$ to DR at zero $\Omega$, which is also known as the Born rate. As expected, for zero rotation the DR matches with the known Born rate which is reflected by the black line. For non-zero rotation, there is an enhancement of DR as compared to the Born rate. With the increase of $\Omega$ the enhancement gets bigger and appears for higher values of $M/T$. At the low invariant mass region, this enhancement is $\sim 15-20\%$. The nature of this figure can be understood from Fig.~\ref{fig:sf_com_diffOmega}, since the DR is calculated using the SF.

\begin{figure}
\includegraphics[scale=0.5]{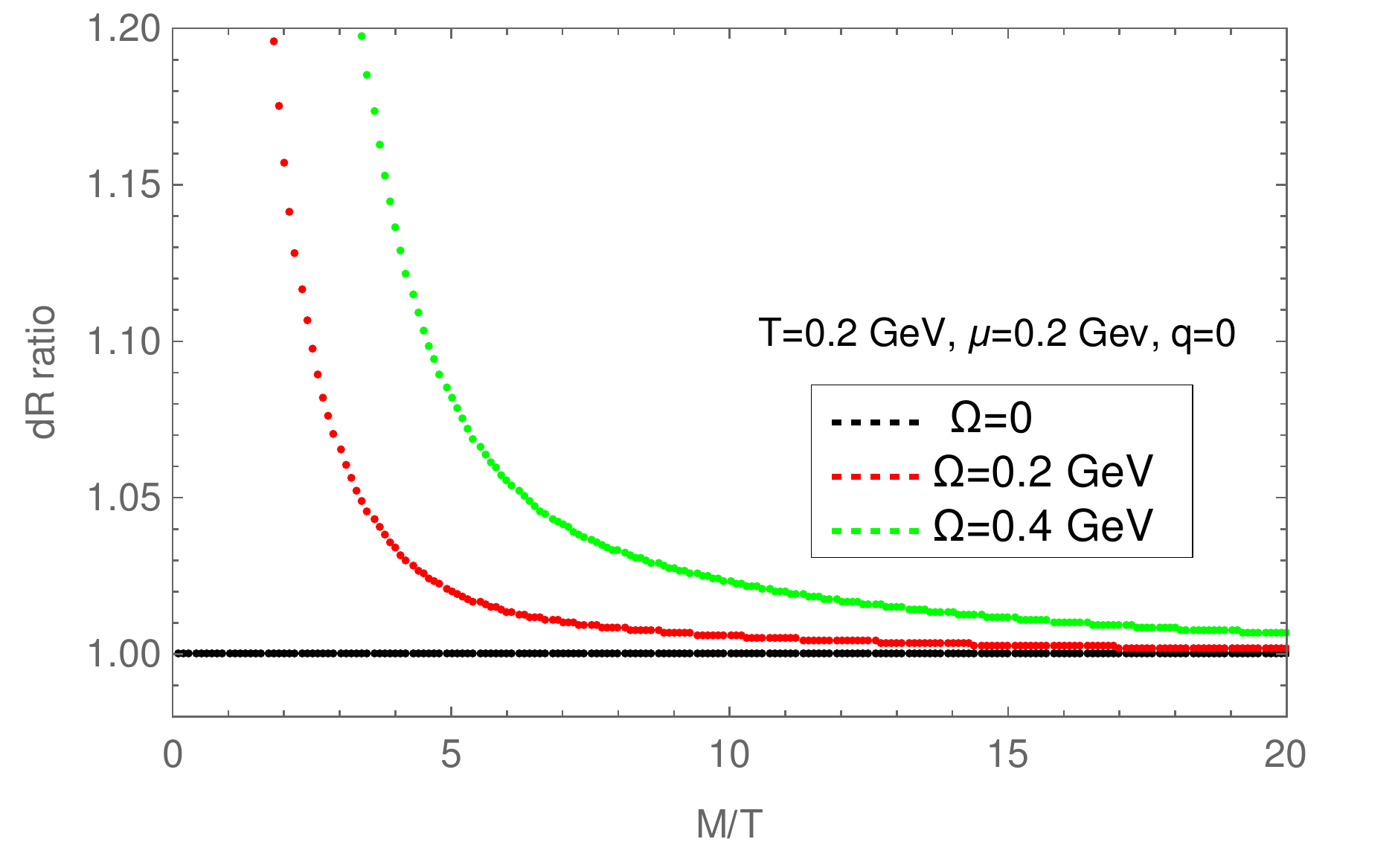}
\caption{Ratios of dilepton rate as a function of temperature scaled invariant mass for different values of angular velocity, with nonzero $T$ and $\mu$.}
\label{fig:dR_ratio_diffOmega_nzMu}
\end{figure}

In Fig.~\ref{fig:dR_ratio_diffOmega_nzMu} we investigate the effects of both the chemical potential and rotation on the DR. This is particularly interesting because of the contrasting effects of the two agents on the SF as evident from  Fig.~\ref{fig:sf_com_diffOmega} and the right panel of Fig.~\ref{fig:sf_com_diffTMu}. The angular velocity enhances the SF whereas the chemical potential diminishes it. It is obtained for three different values of $\Omega$, the values of which along with the other parameters are the same as given in  Fig.~\ref{fig:sf_com_diffOmega_nzMu}. As expected, the zero rotation rate matches with the Born rate and is represented by the black line. With the increase of $\Omega$, there is an enhancement of DR as compared to the Born rate for non-zero chemical potential.
\subsection{Debye mass}
\label{ssec:res_dm}
The Debye mass is related to the temporal part of the CF in the static limit through the relation~\cite{Bandyopadhyay:2016fyd},
\begin{align}
m_D^2=\Pi_{00}(\omega=0,\vert\vec{q}\vert\rightarrow 0).    
\end{align}
The important thing to note here is the order of the limit. First we have to take $\omega=0$ and then take $\vert\vec{q}\vert\rightarrow0$, which is called the static limit\footnote{The same limit taken in reverse order is known as isotropic limit.}. The $00$ component of the CF is given in Eq.~\ref{eq:pi_00}. Taking the static limit in that equation we obtain,
\begin{align}
\Pi^{00}(\omega=0,\vert\vec{q}\vert\rightarrow 0)&= N_{f}N_{c}\sum_{\eta=\pm 1}\int \frac{d^{3}\vec{p}}{(2\pi)^{3}}\frac{1}{T}\left\{\frac{1}{\left(e^{\frac{\sqrt{p^2+m^2}}{T}}+e^{\frac{\eta\Omega}{2T}+\frac{\mu}{T}}\right)} +\frac{1}{\left(e^{\frac{\sqrt{p^2+m^2}}{T}}+e^{\frac{\eta\Omega}{2T}-\frac{\mu}{T}}\right)}\right\}. \label{eq:pi_00}
\end{align}

\begin{figure}
\includegraphics[scale=0.5]{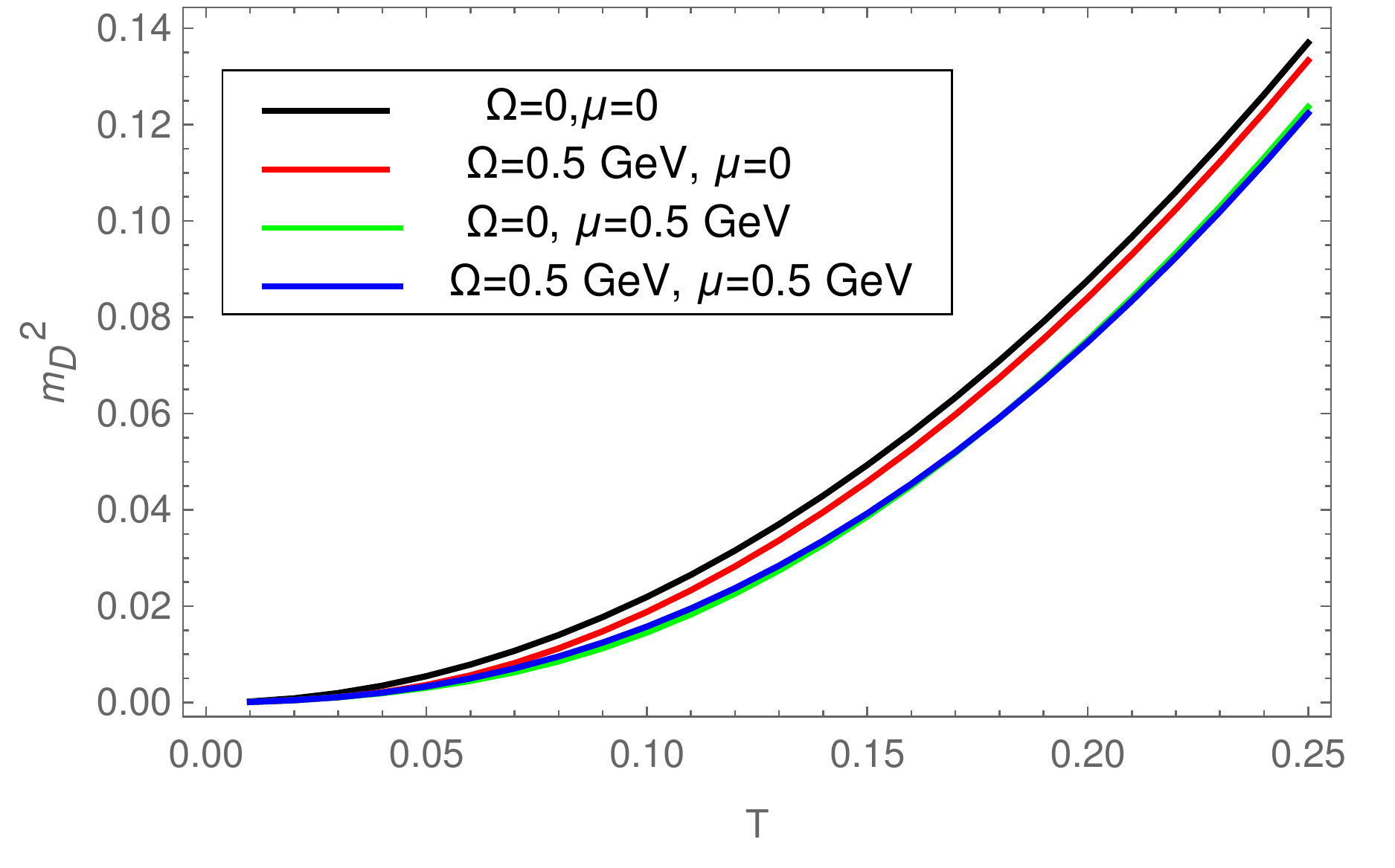}
\caption{Plot of Debye mass as a function of temperature.}
\label{fig:dR_ratio_diffOmega}
\end{figure}
In Fig.~\ref{fig:dR_ratio_diffOmega} we have shown the plot of the Debye mass as a function of temperature for different values of other parameters. The black solid line is obtained when both the angular velocity and the chemical potential are zero. This increasing behaviour of Debye mass as a function of temperature is known from previous studies. 

As we take nonzero values of chemical potential (here $\mu=0.5$ GeV) for zero angular velocity the Debye mass is suppressed as compared to that for zero $\mu$. This is displayed in the figure using a green line. Similar effect is observed for nonzero values of angular velocity with zero $\mu$, we obtain a suppression in the Debye mass, which is shown with a red line in the figure. Though it is to be noted that the suppressing strength for $\Omega$ is less than for similar values of $\mu$. The high value of $\Omega$ that we have used $(0.5)$ GeV in the figure is for demonstration purpose, as such strength of $\Omega$ is highly unlikely to be achieved in HICs. 

We observed that both $\mu$ and $\Omega$ reduce the values of Debye mass when applied to the system separately. In the presence of both of them, the effect of $\Omega$ becomes almost negligible and almost coincide with the line with only nonzero $\mu$. The blue line in the figure represents the case for nonzero values of both $\mu$ and $\Omega$, which almost coincides with the green line. 

\subsection{Ellipticity of dilepton production}

\begin{figure}
\includegraphics[height=5cm]{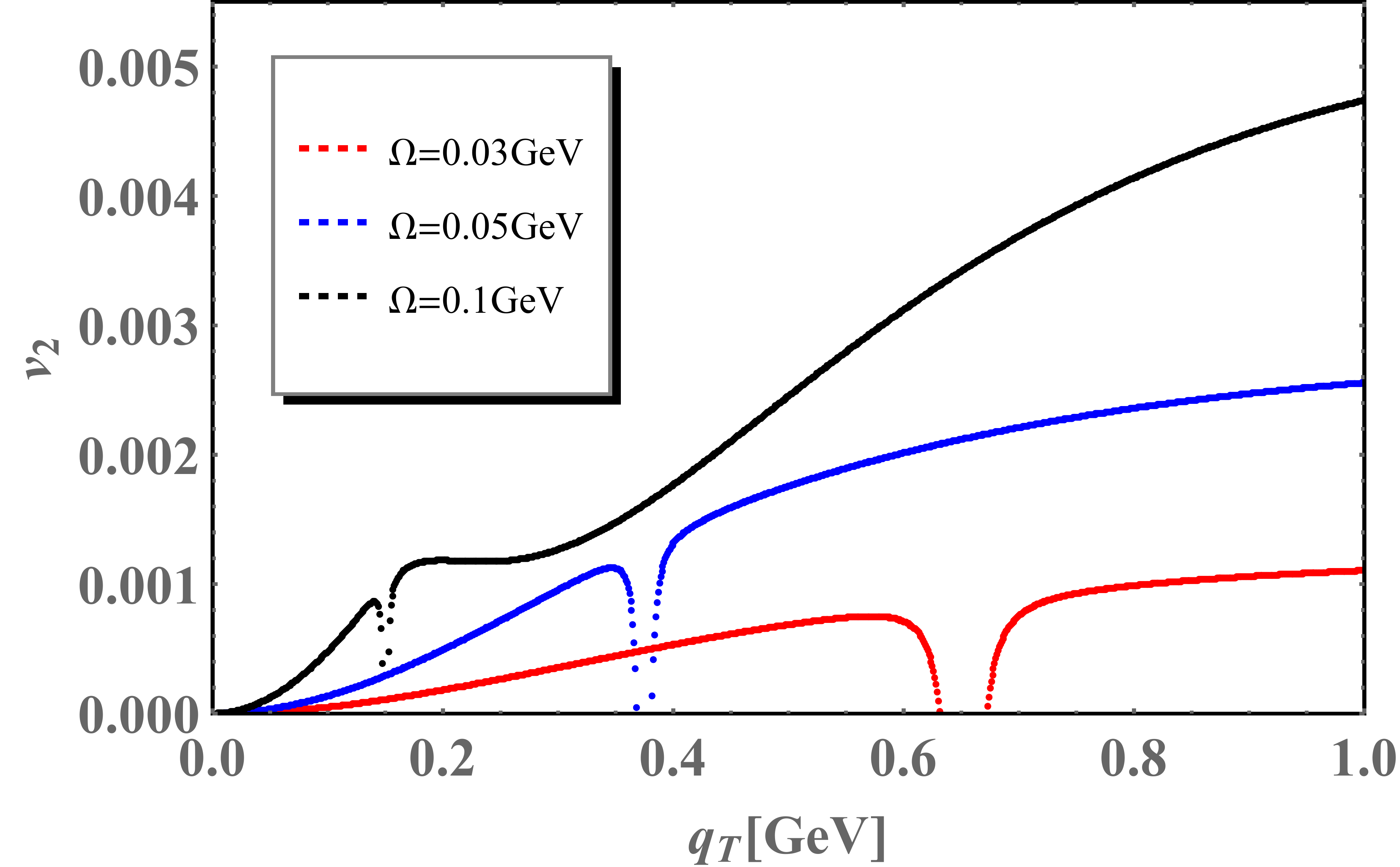}
\includegraphics[height=5cm]{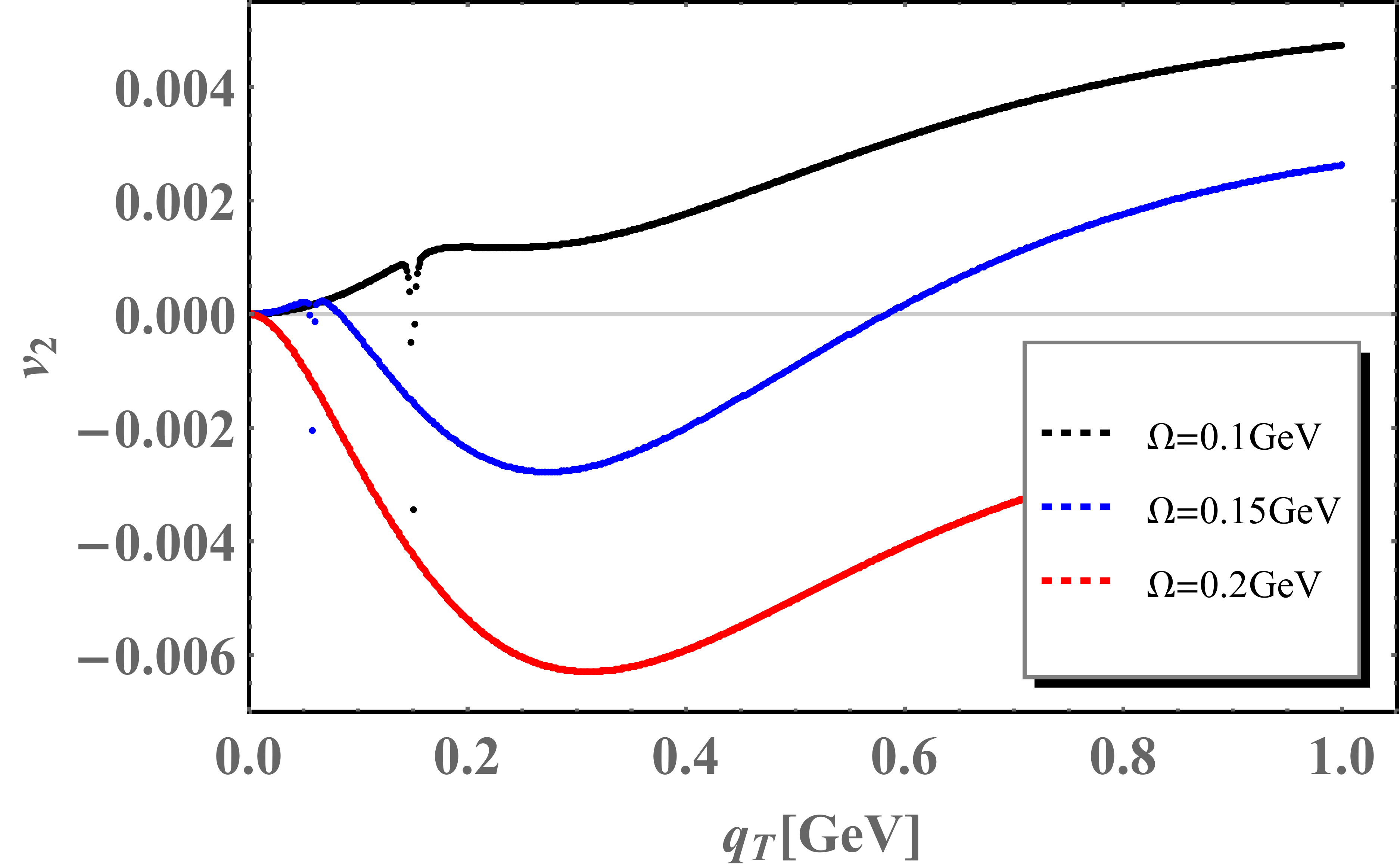}
\caption{Elliptic flow coefficient $v_{2}$ as a function of transverse momentum $q_{T}$ when temperature T=160 MeV and invariant mass M=200MeV. In the left panel, angular velocity $\Omega=0.03,0.05,0.1$ GeV. In the right panel, $\Omega=0.1,0.15,0.2$ GeV}
\label{fig:v2_M200}
\end{figure}

In this section, we show the numerical results for the ellipticity of dilepton production.  Fig.~\ref{fig:v2_M200} shows the elliptic flow $v_{2}$ evolves as a function of transverse momentum $q_{T}$ in different angular velocities. It is observed that rotation plays a complicated role in elliptic flow in different angular velocity regions.

In small angular velocity region, $\Omega=0.03-0.1$ GeV, which is the typical global QGP angular velocity in HIC, the left panel of Fig.~\ref{fig:v2_M200} shows that $v_{2}$ increases with the transverse momentum $q_{T}$ almost monotonically. In this case, positive $v_{2}$ indicates that more dilepton pairs tend to emit from the reaction plane. It is similar to the classical centrifugal effect. Rotating liquid drops will be deformed and accumulate in their equator. Furthermore, a monotonic increase of elliptic flow is in accord with classical centrifugal effect as well, i.e. dilepton pairs with larger momentum $q_{T}$ are affected by larger effective centrifugal force.

It should be noted that singularities show up in $v_{2}$ as a function of $q_{T}$. These singularities emerge from a 4-momentum non-conservation in the rotating frame, because $\delta$ function in the  imaginary part has been modified by $\delta(\omega-E_{p}-E_{k}+\eta\Omega)$, the details of which can be found in Appendix.(\ref{appendix:c}). As a consequence, $p^{\Omega}_{\pm}$ in Subsection (\ref{ssec:com_cf_im}) will reach an extremely large value to satisfy $\omega-E_{p}-E_{k}+\eta\Omega=0$. We have shown that the rotation effect had modified the domain of integration by  $p^{\Omega}_{\pm}=\pm\frac{q}{2}+\frac{\omega+\eta\Omega}{2}\sqrt{1-\frac{4M_{f}^2}{(\omega+\eta\Omega)^2-q^{2}}}$. Resonance is generated at $q_{t}=\omega+\eta\Omega$, and the singularity will show up with uniform rotation, which is similar to the forced oscillator without damping. However, in HICs, the global rotation will decrease as the QGP evolves~\cite{Jiang:2016woz}. In this case, the divergence will not show up in the realistic system. Currently, theoretical analysis is unable to adopt all the behavior of a rotating QGP in a hydrodynamic simulation. 

Though hydrodynamic simulations indicate that global rotation of QGP is hard to reach the magnitude of 0.1 GeV, the effect of a larger magnitude of local vorticity on quark matter still attracts theoretical studies. In the right panel of Fig.~\ref{fig:v2_M200}, angular velocities range from 0.1 GeV to 0.2GeV. For angular velocity $\Omega=0.15$ GeV, $v_{2}$ firstly increases a little bit then decreases quickly to negative values at transverse momentum $q_{T}\sim 0.08$ GeV. At $q_{T}\sim 0.27$ GeV, negative elliptic flow reaches a minimal value $v_{2,min}\sim -0.0028$. Finally, $v_{2}$ increases and turns to positive at $q_{T}\sim 0.58$ GeV. For $\Omega=0.2$ GeV, elliptic flow $v_{2}$ is negative at small transverse momentum region and it reaches its minimum $v_{2,min}\sim -0.0063$ at $q_{T}\sim 0.31$ GeV,  eventually $v_{2}$ increases but still keeps negative at larger transverse momentum. The negative $v_{2}$ can be understood from the following argument. Lepton pairs tend to emit from the direction of the rotating axis, which is the direction of quark polarization as well. It is reasonable that larger angular velocities create more dileptons along the rotating axis, because large vorticity creates quark polarization phenomenon and generates polarization of final particles~\cite{STAR:2017ckg}. For example, polarized cross section $\frac{d\sigma}{d\Omega}(q_{L}\bar{q}_{R}\rightarrow l \bar{l})\sim 1+\cos^{2}{\theta^{*}}$ contributes an elliptical distribution with $\theta^{*}$ the angle between the polarization direction and the emission direction. As a consequence, in polarized QGP, lepton pairs tend to emit from $\theta^{*}=0,\pi$ direction, and lower negative $v_{2}$ will emerge in a more polarized medium. In this opinion, the non-monotonic behavior of $v_{2}$ can be explained.

The non-monotonic behavior of elliptic flow reflects that two rotational effects, the centrifugal effect and the spin polarization effect, compete with each other. The centrifugal effect dominates in small angular velocity and large $q_{T}$ region, and the quark spin polarization is more significant in the larger angular velocity region. Consequently, a convex down behavior shows up at $q_{T}\sim 0.3$ GeV for large angular velocity, the larger the angular velocity the stronger the convexity will be.

Finally, the magnitude of elliptic flow is around $\mathcal{O}(v_{2})\sim 10^{-3}-10^{-2}$. It is small but not a negligible contribution from polarization function under rotation, because the contribution is sufficiently large in comparison with $v_{2}$ from hydrodynamic initial conditions. For example, fluctuating “hot spots” contributes an elliptic flow $v_{2}$ around 0.02~\cite{Xu:2014ada}, and an initial condition for an anisotropic QGP contribute a $v_2$ around $0.01\sim 0.05$~\cite{Kasmaei:2018oag}.

\section{Conclusion}
\label{sec:con}
In non-central heavy-ion collisions (HIC), the consideration of rotation in different QCD related phenomena becomes crucial, and large vorticity becomes inevitable for the study of a hot and dense quark-gluon plasma (QGP) created in such collisions. In this article, we investigate the rotation effect on dilepton pairs production by modifying the photon polarization tensor in presence of a hot and dense QCD medium subjected to a non-zero angular velocity.

In analytical calculation, the photon polarization tensor is analogous to the vector current-current correlation function (CF). The CF and its spectral representation are reliable tools to explore many-particle systems. It is particularly useful that the dilepton rate (DR) can be directly estimated from such spectral representation of CF. In this work, we calculate the spectral function (SF) and the DR for a hot and dense QGP medium exposed to a strong rotation. It is found that the rotation affects the polarization function beyond effective chemical potential, while the angular velocity plays a similar role to the effective chemical potential in  QCD phase diagram~\cite{Jiang:2016wvv,Wang:2018sur}. The angular velocity terms not only appear in distribution functions, but also induce an anisotropy of polarization function and its imaginary part, which indicates corresponding anisotropy of dilepton rate. 

Numerical results show that the SF gets enhanced in the presence of a non-zero angular velocity as compared to the zero case when plotted as a function of the invariant mass scaled with the temperature. This enhancement is remarkably large in a lower invariant mass region where thermal dilepton pairs are dominant sources. However, the enhancement effect of rotation will be reduced by the chemical potential in HIC. This is obtained while the other parameters like temperature, chemical potential and external momentum are kept at fixed values. The intertwining effects of the rotation with chemical potential on the SF are explored while keeping the strength of the rotation fixed to a non-zero value. In such cases, the qualitative behavior of the SF remains the same as in the case of zero rotation. Our observation of enhancement in the SF with the increase of the rotational strength could indeed be an interesting finding for the HIC related phenomenology, particularly properties related to the SF. 

Then the DR can be easily calculated from the SF. The lepton pairs can move freely inside the QCD medium because of their long mean free path, carrying information of the stages at which they are produced to the detector. Thus, their detection can help us understand different stages of HICs. As a spectral property, the DR coming out of a rotating QGP must be influenced by the angular velocity. The enhancement of SF with the increasing angular velocity also gets reflected in the lepton pair production. When the rate is plotted as a ratio to the Born rate we observe an increase of $15-20\%$ at the low invariant mass region. This, in our opinion, is an interesting observation which states the importance of consideration of rotational effects in relevant HIC related exercises.

In addition, the Debye mass has also been evaluated from the temporal component of the polarization function. For the temporal component, the angular velocity $\Omega$ merely acts as an effective chemical potential and appears only in the distribution functions. Consequently, Debye mass, which signifies the electromagnetic screening, behaves quite similarly in finite $\mu$ and $\Omega$ when plotted as a function of temperature. In other words, the effect of $\Omega$ on the Debye mass is to suppress it just like the impact of non-zero chemical potential.

The most significant result of this investigation is the azimuthal anisotropy of the dilepton production induced by rotation. In the low angular velocity region, the elliptic flow $v_{2}$ increases from zero to positive value, which indicates that lepton pairs tend to emit from the reaction plane. In this case, the rotation affects DR as a centrifugal effect. In a relatively large magnitude of $\Omega=0.1-0.2$ GeV, though larger angular velocities indicate more remarkable centrifugal effects, a competitive spin polarization effect induced by rotation drives $v_{2}$ decreasing to a negative minimum which indicates an emission preference for $\pm z$ direction. The competition between the centrifugal effect and the spin polarization effect results in a convex down behavior of $v_{2}$ as a function of $q_{T}$ in relatively large magnitude of angular velocity. It is noticed that quark spin polarization induces a negative $v_2$ in the case of large angular velocity.

Here, we want to make some final comments regarding similar calculations done in presence of magnetic fields. In the introduction, we drew an analogy between the rotation and the magnetic fields. We talked about different analogous phenomena which can be observed in presence of either of the two. The calculation of DR in presence of a magnetic field also sees enhancement both in the case of lowest Landau level approximation~\cite{Bandyopadhyay:2016fyd} and in arbitrary strength of the fields~\cite{Das:2021fma}. Thus, it presents us with a complex and interesting possibility of exploring the rate in presence of both rotation and magnetic field, which we plan to explore in the near future.  

\begin{acknowledgements}
We thank Wei Chen, Kun Xu and Xinyang Wang for their useful discussion. M.H.is supported by the NSFC under Grant Nos. 11725523 and 11735007,  Chinese Academy of Sciences under Grant No. XDPB09 and XDPB15, the start-up funding from the University of Chinese Academy of Sciences (UCAS), and the Fundamental Research Funds for the Central Universities. CAI was supported by the Chinese Academy of Sciences President’s International Fellowship Initiative under Grant No. 2020PM0064. 
\end{acknowledgements}

\appendix
\section{Components of the polarisation function under rotation}
 \label{appendix:a}
 The polarisation function with one loop contribution can be expressed as
\begin{equation}
\Pi^{ab}(q)=-i \int d^{4}\tilde{r}Tr_{sfc}[i \gamma^{a}S(0;\tilde{r})i \gamma^{b}S(\tilde{r};0)]e^{i q\cdot \tilde{r}}
\end{equation}
where $S(\tilde{r};\tilde{r'})$ is propagator in rotating medium, its explicit form was shown in Eq.(\ref{eq:quark_prop}). Diagonal components can be calculated as following Ref.(\cite{Wei:2020xfd}). Now, we use $\Pi^{aa}(q)$ for respective components for polarisation functions. $\Pi^{a}_{a}(q)$ stands for the summation for components. Rotation will make a difference of the denominator of each $\Pi^{aa}(q)$, so the numerators can not be added up straightforwardly. 

For diagonal component:
\begin{equation}
\begin{aligned}
\Pi^{00}(q)&=
-2N_{f}N_{c}\sum_{\eta=\pm 1}\int \frac{d^{4} p}{(2\pi)^{4}} 
\frac{M_{f}^2+\left(p_{0}+\frac{\eta\Omega}{2}\right) \left(p_{0}+q_{0}+\frac{\eta\Omega}{2}\right)+ \vec{p}\cdot(\vec{p}+\vec{q})}
{\left[\left(p_{0}+\frac{\eta\Omega }{2}\right)^2-\vec{p}^2-M_{f}^2\right] \left[\left(p_{0}+q_{0}+\frac{\eta\Omega }{2}\right)^2-(\vec{p}+\vec{q})^2-M_{f}^2\right]}\\
\Pi^{11}(q)&=
-2N_{f}N_{c}\sum_{\eta=\pm 1}\int \frac{d^{4} p}{(2\pi)^{4}} 
\frac{M_{f}^2-\left(p_{0}+\frac{\eta\Omega}{2}\right) \left(p_{0}+q_{0}-\frac{\eta\Omega}{2}\right)-2 p_{x}(p_{x}+q_{x})+ \vec{p}\cdot(\vec{p}+\vec{q})}
{\left[\left(p_{0}+\frac{\eta\Omega }{2}\right)^2-\vec{p}^2-M_{f}^2\right] \left[\left(p_{0}+q_{0}-\frac{\eta\Omega }{2}\right)^2-(\vec{p}+\vec{q})^2-M_{f}^2\right]}\\
\Pi^{22}(q)&=
-2N_{f}N_{c}\sum_{\eta=\pm 1}\int \frac{d^{4} p}{(2\pi)^{4}} 
\frac{M_{f}^2-\left(p_{0}+\frac{\eta\Omega}{2}\right) \left(p_{0}+q_{0}-\frac{\eta\Omega}{2}\right)-2 p_{y}(p_{y}+q_{y})+ \vec{p}\cdot(\vec{p}+\vec{q})}
{\left[\left(p_{0}+\frac{\eta\Omega }{2}\right)^2-\vec{p}^2-M_{f}^2\right] \left[\left(p_{0}+q_{0}-\frac{\eta\Omega }{2}\right)^2-(\vec{p}+\vec{q})^2-M_{f}^2\right]}\\
\Pi^{33}(q)&=
-2N_{f}N_{c}\sum_{\eta=\pm 1}\int \frac{d^{4} p}{(2\pi)^{4}} 
\frac{M_{f}^2- \left(p_{0}+\frac{\eta\Omega}{2}\right) \left(p_{0}+q_{0}+\frac{\eta\Omega}{2}\right)-2 p_{z}(p_{z}+q_{z})+ \vec{p}\cdot(\vec{p}+\vec{q})}
{\left[\left(p_{0}+\frac{\eta\Omega }{2}\right)^2-\vec{p}^2-M_{f}^2\right] \left[\left(p_{0}+q_{0}+\frac{\eta\Omega }{2}\right)^2-(\vec{p}+\vec{q})^2-M_{f}^2\right]}
\end{aligned}
\end{equation}
Due to symmetric analysis for integration, $\Pi^{11}(q)$ and $\Pi^{22}(q)$ will be equal.
This is at zero temperature. To go to nonzero temperature, one needs to make the following transformations,
\begin{equation}
p_{0}\rightarrow i \tilde{\omega}_{N},\hspace{10pt}
q_{0}\rightarrow i \nu_{n},\hspace{10pt}
\int \frac{p_{0}}{2\pi}\rightarrow i T\sum_{N},\hspace{10pt}
\tilde{\omega}_{N}=(2N+1)\pi T.
\end{equation}
Here also we can write down the components as well,
\begin{equation}
\Pi^{00}(i\nu_{n},\vec{q})=2N_{f}N_{c}T\sum_{N}\sum_{\eta=\pm 1}\int\frac{d^3\vec{p}}{(2\pi)^3}\frac{(i\tilde{\omega}_{N}+i\nu_{n}+\frac{1}{2}\eta\Omega+\mu)(\tilde{\omega}_{N}+\frac{1}{2}\eta\Omega+\mu)+M^2+(\vec{p}+\vec{q})\cdot\vec{q}}
{\left[(i\tilde{\omega}_{N}+i\nu_{n}+\frac{1}{2}\eta\Omega+\mu)^2-(\vec{p}+\vec{q})^2-M_{f}^2\right]\left[(i\tilde{\omega}_{N}+\frac{1}{2}\eta\Omega+\mu)^2-\vec{p}^2-M^2\right]}
\end{equation}

\begin{equation}
\begin{aligned}
\Pi^{11}(i\nu_{n},\vec{q})&=
2N_{f}N_{c}T\sum_{N}\sum_{\eta=\pm 1}\int \frac{d^{3}\vec{p}}{(2\pi)^{3}} \\
&-\frac{ M_{f}^2-\left(i \tilde{\omega}_{N}+\mu+\frac{\eta\Omega}{2}\right) \left(i \tilde{\omega}_{N}+\mu+i\nu_{n}-\frac{\eta\Omega}{2}\right)- p_{x}(p_{x}+q_{x})+p_{y}(p_{y}+q_{y})+p_{z}(p_{z}+q_{z})}
{\left[\left(i \tilde{\omega}_{N}+\mu+\frac{\eta\Omega }{2}\right)^2-\vec{p}^2-M_{f}^2\right] \left[\left(i \tilde{\omega}_{N}+\mu+i\nu_{n}-\frac{\eta\Omega }{2}\right)^2-(\vec{p}+\vec{q})^2-M_{f}^2\right]}
\end{aligned}
\end{equation}

\begin{equation}
 \begin{aligned}
\Pi^{22}(i\nu_{n},\vec{q})&=
2N_{f}N_{c}T\sum_{N}\sum_{\eta=\pm 1}\int \frac{d^{3}\vec{p}}{(2\pi)^{3}} \\
&-\frac{M_{f}^2-\left(i \tilde{\omega}_{N}+\mu-\frac{\eta\Omega}{2}\right) \left(i \tilde{\omega}_{N}+\mu+i\nu_{n}+\frac{\eta\Omega}{2}\right)+ p_{x}(p_{x}+q_{x})-p_{y}(p_{y}+q_{y})+p_{z}(p_{z}+q_{z})}
{\left[\left(i \tilde{\omega}_{N}+\mu-\frac{\eta\Omega }{2}\right)^2-\vec{p}^2-M_{f}^2\right] \left[\left(i \tilde{\omega}_{N}+\mu+i\nu_{n}+\frac{\eta\Omega }{2}\right)^2-(\vec{p}+\vec{q})^2-M_{f}^2\right]}
\end{aligned}
\end{equation}

\begin{equation}
\begin{aligned}
\Pi^{33}(i\nu_{n},\vec{q})&=
2N_{f}N_{c}T\sum_{N}\sum_{\eta=\pm 1}\int \frac{d^{3}\vec{p}}{(2\pi)^{3}} \\
&-\frac{M_{f}^2-\left(i \tilde{\omega}_{N}+\mu+\frac{\eta\Omega}{2}\right) \left(i \tilde{\omega}_{N}+\mu+i\nu_{n}+\frac{\eta\Omega}{2}\right)+p_{x}(p_{x}+q_{x})+p_{y}(p_{y}+q_{y})-p_{z}(p_{z}+q_{z})}
{\left[\left(i \tilde{\omega}_{N}+\mu+\frac{\eta\Omega }{2}\right)^2-\vec{p}^2-M_{f}^2\right] \left[\left(i \tilde{\omega}_{N}+\mu+i\nu_{n}+\frac{\eta\Omega }{2}\right)^2-(\vec{p}+\vec{q})^2-M_{f}^2\right]}
\end{aligned}
\end{equation}

\section{Matsubara Summation calculation}
\label{appendix:b}
Now we give an example for Matsubara Summation calculation with angular velocity $\Omega$:
\begin{equation}
\begin{aligned}
&\sum_{N}\frac{1}{i \tilde{\omega}_{N}+\omega+\mu-\frac{\Omega}{2}-E_{k}}\cdot\frac{1}{i \tilde{\omega}_{N}+\mu+\frac{\Omega}{2}-E_{p}}\\
=&\sum_{N}\frac{-i}{ (2N+1)\pi T+i(-\omega-\mu+\frac{\Omega}{2}+E_{k})}\cdot\frac{-i}{(2N+1)\pi T+i(-\mu-\frac{\Omega}{2}+E_{p})}\\
=&-\frac{\coth\frac{1}{2T}(E_{k}-\omega-\mu+\frac{\Omega}{2}-i\pi T)-\coth\frac{1}{2T}(E_{p}-\mu-\frac{\Omega}{2}-i\pi T)}{2 T[(E_{k}-\omega-\mu+\frac{\Omega}{2}-i\pi T)-(E_{p}-\mu-\frac{\Omega}{2}-i\pi T)]}\\
=&\frac{1}{T}\cdot\frac{1}{\omega+E_{p}-E_{k}-\Omega}[f(E_{p}-\mu-\frac{\Omega}{2})-f(E_{k}-\mu-\omega+\frac{\Omega}{2})]
\end{aligned}
\end{equation}
Here, we have applied:
\begin{equation}
\sum_{N}\frac{1}{N+i x}\frac{1}{N+i y}=\frac{\pi}{x-y}[\coth(\pi x)-\coth(\pi y)]
\end{equation}
We can see that a $\Omega$ term in denominator. Similarly, we have
\begin{equation}
\begin{aligned}
\sum_{N}\frac{1}{i \tilde{\omega}_{N}+\omega+\mu-\frac{\Omega}{2}-E_{k}}\cdot\frac{1}{i \tilde{\omega}_{N}+\mu+\frac{\Omega}{2}+E_{p}}=&\frac{1}{T}\cdot\frac{1}{\omega-E_{p}-E_{k}-\Omega}[1-f(E_{p}+\mu+\frac{\Omega}{2})-f(E_{k}-\mu-\omega+\frac{\Omega}{2})]
\\
\sum_{N}\frac{1}{i \tilde{\omega}_{N}+\omega+\mu-\frac{\Omega}{2}+E_{k}}\cdot\frac{1}{i \tilde{\omega}_{N}+\mu+\frac{\Omega}{2}-E_{p}}
=&\frac{1}{T}\cdot\frac{1}{\omega+E_{p}+E_{k}-\Omega}[-1+f(E_{p}-\mu-\frac{\Omega}{2})+f(E_{k}+\mu+\omega-\frac{\Omega}{2})]
\\
\sum_{N}\frac{1}{i \tilde{\omega}_{N}+\omega+\mu-\frac{\Omega}{2}+E_{k}}\cdot\frac{1}{i \tilde{\omega}_{N}+\mu+\frac{\Omega}{2}+E_{p}}
=&\frac{1}{T}\cdot\frac{1}{\omega-E_{p}+E_{k}-\Omega}[-f(E_{p}+\mu+\frac{\Omega}{2})+f(E_{k}+\mu+\omega-\frac{\Omega}{2})]
\end{aligned}
\end{equation}
\section{The imaginary part of polarisation function under rotation}
\label{appendix:c}
In Subsection(\ref{ssec:com_cf}), whole polarisation functions have been presented. Imaginary part of polarisation function can be extracted by:
\begin{equation}
Im\frac{1}{x+i \epsilon}=-\pi \delta(x)
\end{equation}
We write down all the components here. The temporal part is:
\begin{equation}
\begin{aligned}
Im\Pi^{00}(\omega,\vec{q})
&=-\frac{\pi}{2} N_{f}N_{c}\sum_{\eta=\pm 1}\int \frac{d^{3}\vec{p}}{(2\pi)^{3}} \\
&\times\frac{1}{E_{p}E_{k}}\left\{[E_{p}E_{k}+\vec{p}\cdot\vec{k}+M_{f}^{2}]\right.\\
&[f(E_{p}-\mu-\frac{\eta\Omega}{2})+f(E_{p}+\mu-\frac{\eta\Omega}{2})][\delta(\omega+E_{p}-E_{k})-\delta(\omega-E_{p}+E_{k})]\\
&+[E_{p}E_{k}-\vec{p}\cdot\vec{k}-M_{f}^{2}]\delta(\omega-E_{p}-E_{k})\\
&\times\left.[1-f(E_{p}-\mu-\frac{\eta\Omega}{2})-f(E_{p}+\mu-\frac{\eta\Omega}{2})]\right\}
\end{aligned}
\end{equation}
The transverse part is:
\begin{equation}
\label{ImPi1122}
\begin{aligned}
&Im[\Pi^{11}(\omega,\vec{q})+\Pi^{22}(\omega,\vec{q})]\\
&=-\frac{1}{2}\pi N_{f}N_{c}\sum_{\eta=\pm 1}\int \frac{d^{3}\vec{p}}{(2\pi)^{3}}
\left\{\frac{2E_{p}E_{k}-2 p_{z}(p_{z}+q_{z})-2M_{f}^{2}}{E_{p}E_{k}}\right.\\
&\times[f(E_{k}-\mu-\frac{\eta\Omega}{2})-f(E_{p}-\mu+\frac{\eta\Omega}{2})-f(E_{p}+\mu+\frac{\eta\Omega}{2})+f(E_{k}+\mu-\frac{\eta\Omega}{2})]\delta(\omega-E_{p}+E_{k}-\eta\Omega)\\
&+\frac{2E_{p}E_{k}+2 p_{z}(p_{z}+q_{z})+2M_{f}^{2}}{E_{p}E_{k}}\left.[1-f(E_{p}-\mu+\frac{\eta\Omega}{2})-f(E_{k}+\mu+\frac{\eta\Omega}{2})]\delta(\omega-E_{p}-E_{k}-\eta\Omega)\right\}
\end{aligned}
\end{equation}
The longitudinal part is:
\begin{equation}
\begin{aligned}
Im\Pi^{33}(\omega,\vec{q})
&=-\frac{\pi}{2} N_{f}N_{c}\sum_{\eta=\pm 1}\int \frac{d^{3}\vec{p}}{(2\pi)^{3}} \\
&\times\frac{1}{E_{p}E_{k}}\left\{[E_{p}E_{k}-[+p_{x}(p_{x}+q_{x})+ p_{y}(p_{y}+q_{y})- p_{z}(p_{z}+q_{z})]-M_{f}^{2}]\right.\\
&\times[f(E_{p}-\mu-\frac{\eta\Omega}{2})+f(E_{p}+\mu-\frac{\eta\Omega}{2})][\delta(\omega-E_{k}+E_{p})-\delta(\omega-E_{p}+E_{k})]\\
&+[E_{p}E_{k}+[p_{x}(p_{x}+q_{x})+ p_{y}(p_{y}+q_{y})- p_{z}(p_{z}+q_{z})]+M_{f}^{2}]\delta(\omega-E_{p}-E_{k})\\
&\times\left.[1-f(E_{p}-\mu-\frac{\eta\Omega}{2})-f(E_{p}+\mu-\frac{\eta\Omega}{2})]\right\}
\end{aligned}
\end{equation}
Now, we have to deal with the $\delta$ functions and integration in Eq.(\ref{ImPi1122}). Because branch cuts are still confusing, we consider the $\delta$ function of $\delta(\omega-E_{p}-E_{k}+\eta\Omega)$

For simplification, we will interchange variable of $p$ and $k$ and use an polar coordinate whose polar axis is $\vec{q}$-axis. For calculation, we will choose a polar coordinate, and $\vec{q}$ will be the polar axis. The new polar axis $\vec{q}$ has an angle $\alpha$ with original $z-$ axis.
\begin{figure}
  \centering
  \includegraphics[width=200pt]{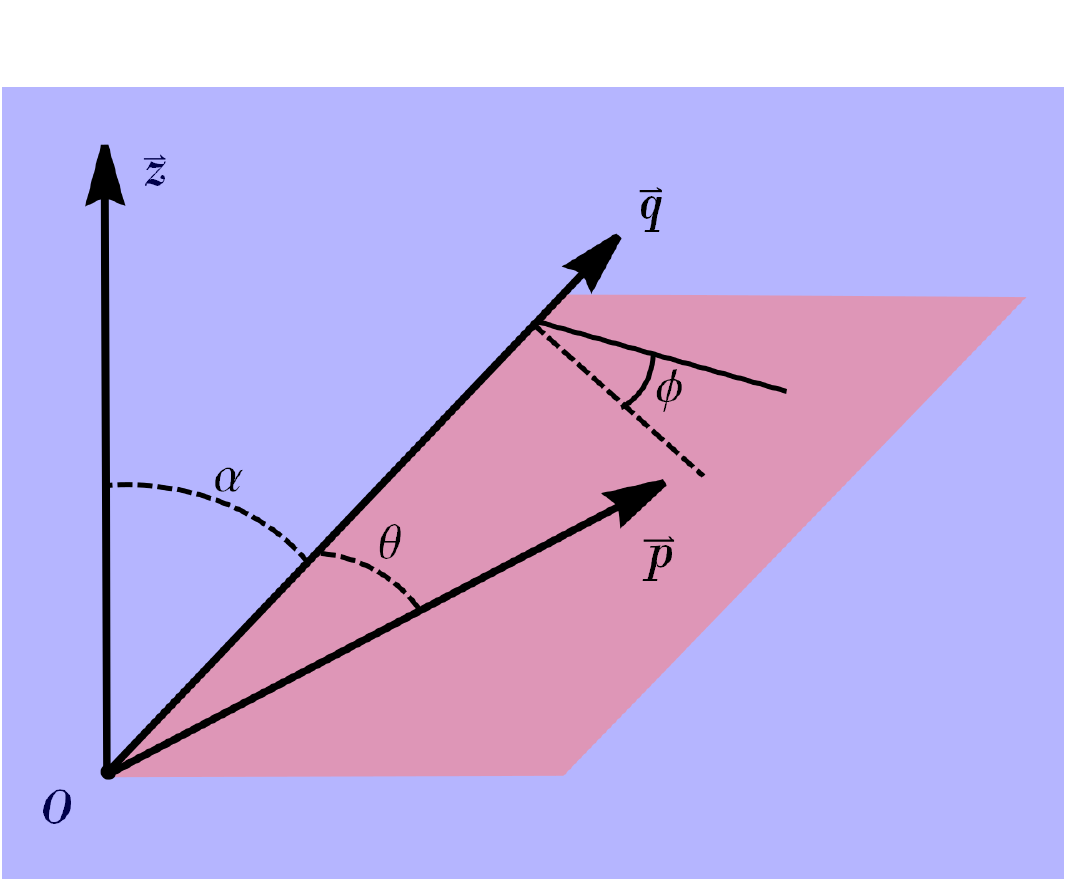}\\
  \caption{The angles for integration.}\label{fig:angles}
\end{figure}

As a consequence, $p_{z}$ and $q_{z}$ can be represented as following:
\begin{equation}
\left\{
  \begin{aligned}
    q_{z} &=q \cos\alpha \\
    p_{z} &=p(\cos\alpha\cos\theta-\sin\alpha\sin\theta\cos\phi)
  \end{aligned}
\right.
\end{equation}
where $\theta$ and $\phi$ are coordinates for $\vec{p}$. Relevant angles and vectors are shown in Fig.~\ref{fig:angles}. Eventually, $\vec{p}$ will be integrated, so transverse part can be expressed as a function of $q$ and $q_{z}$: 
\begin{equation}
\begin{aligned}
&Im[\Pi^{11}(\omega,\vec{q})+\Pi^{22}(\omega,\vec{q})]\\
&=-\frac{1}{2}\pi N_{f}N_{c}\sum_{\eta=\pm 1}\int_{p^{\Omega}_{-}}^{p^{\Omega}_{+}} \frac{p d p}{(2\pi)^{2}}\\
&\times\left\{\frac{2 E_{p}(\omega-E_{p}+\eta\Omega)+[(3\frac{q_{z}^2}{q^{2}}-1)(\frac{(\omega+\eta\Omega)^{2}-2(\omega+\eta\Omega)\sqrt{p^{2}+M_{f}^{2}}-q^{2}}{2q})^{2}+p^{2}(1-\frac{q_{z}^{2}}{q^{2}})+2 \frac{q_{z}^{2}}{q}\frac{(\omega+\eta\Omega)^{2}-2(\omega+\eta\Omega)\sqrt{p^{2}+M_{f}^{2}}-q^{2}}{2q}]+2 M_{f}^{2}}{q E_{p}}\right.
\\
&\times\left.[1-f(E_{p}-\mu-\frac{\eta\Omega}{2})-f(E_{p}+\mu-\frac{\eta\Omega}{2})]\right\}
\end{aligned}
\end{equation}
It implies a angular distribution on dilepton production. $Im\Pi^{33}(\omega,\vec{q})$ also can be obtain similarly.

\bibliography{ref}

\begin{thebibliography}{64}%
\makeatletter
\providecommand \@ifxundefined [1]{%
 \@ifx{#1\undefined}
}%
\providecommand \@ifnum [1]{%
 \ifnum #1\expandafter \@firstoftwo
 \else \expandafter \@secondoftwo
 \fi
}%
\providecommand \@ifx [1]{%
 \ifx #1\expandafter \@firstoftwo
 \else \expandafter \@secondoftwo
 \fi
}%
\providecommand \natexlab [1]{#1}%
\providecommand \enquote  [1]{``#1''}%
\providecommand \bibnamefont  [1]{#1}%
\providecommand \bibfnamefont [1]{#1}%
\providecommand \citenamefont [1]{#1}%
\providecommand \href@noop [0]{\@secondoftwo}%
\providecommand \href [0]{\begingroup \@sanitize@url \@href}%
\providecommand \@href[1]{\@@startlink{#1}\@@href}%
\providecommand \@@href[1]{\endgroup#1\@@endlink}%
\providecommand \@sanitize@url [0]{\catcode `\\12\catcode `\$12\catcode
  `\&12\catcode `\#12\catcode `\^12\catcode `\_12\catcode `\%12\relax}%
\providecommand \@@startlink[1]{}%
\providecommand \@@endlink[0]{}%
\providecommand \url  [0]{\begingroup\@sanitize@url \@url }%
\providecommand \@url [1]{\endgroup\@href {#1}{\urlprefix }}%
\providecommand \urlprefix  [0]{URL }%
\providecommand \Eprint [0]{\href }%
\providecommand \doibase [0]{https://doi.org/}%
\providecommand \selectlanguage [0]{\@gobble}%
\providecommand \bibinfo  [0]{\@secondoftwo}%
\providecommand \bibfield  [0]{\@secondoftwo}%
\providecommand \translation [1]{[#1]}%
\providecommand \BibitemOpen [0]{}%
\providecommand \bibitemStop [0]{}%
\providecommand \bibitemNoStop [0]{.\EOS\space}%
\providecommand \EOS [0]{\spacefactor3000\relax}%
\providecommand \BibitemShut  [1]{\csname bibitem#1\endcsname}%
\let\auto@bib@innerbib\@empty
\bibitem [{\citenamefont {Muller}(1985)}]{Muller:1983ed}%
  \BibitemOpen
  \bibfield  {author} {\bibinfo {author} {\bibfnamefont {B.}~\bibnamefont
  {Muller}},\ }\bibfield  {title} {\bibinfo {title} {{THE PHYSICS OF THE QUARK
  - GLUON PLASMA}},\ }\href@noop {} {\bibfield  {journal} {\bibinfo  {journal}
  {Lect. Notes Phys.}\ }\textbf {\bibinfo {volume} {225}},\ \bibinfo {pages}
  {1} (\bibinfo {year} {1985})}\BibitemShut {NoStop}%
\bibitem [{\citenamefont {Heinz}\ and\ \citenamefont
  {Jacob}(2000)}]{Heinz:2000bk}%
  \BibitemOpen
  \bibfield  {author} {\bibinfo {author} {\bibfnamefont {U.~W.}\ \bibnamefont
  {Heinz}}\ and\ \bibinfo {author} {\bibfnamefont {M.}~\bibnamefont {Jacob}},\
  }\bibfield  {title} {\bibinfo {title} {{Evidence for a new state of matter:
  An Assessment of the results from the CERN lead beam program}},\ }\href@noop
  {} {\  (\bibinfo {year} {2000})},\ \Eprint
  {https://arxiv.org/abs/nucl-th/0002042} {arXiv:nucl-th/0002042} \BibitemShut
  {NoStop}%
\bibitem [{\citenamefont {Yagi}\ \emph {et~al.}(2005)\citenamefont {Yagi},
  \citenamefont {Hatsuda},\ and\ \citenamefont {Miake}}]{Yagi:2005yb}%
  \BibitemOpen
  \bibfield  {author} {\bibinfo {author} {\bibfnamefont {K.}~\bibnamefont
  {Yagi}}, \bibinfo {author} {\bibfnamefont {T.}~\bibnamefont {Hatsuda}},\ and\
  \bibinfo {author} {\bibfnamefont {Y.}~\bibnamefont {Miake}},\ }\href@noop {}
  {\emph {\bibinfo {title} {{Quark-gluon plasma: From big bang to little
  bang}}}},\ Vol.~\bibinfo {volume} {23}\ (\bibinfo {year} {2005})\BibitemShut
  {NoStop}%
\bibitem [{\citenamefont {Fukushima}(2012)}]{Fukushima:2011jc}%
  \BibitemOpen
  \bibfield  {author} {\bibinfo {author} {\bibfnamefont {K.}~\bibnamefont
  {Fukushima}},\ }\bibfield  {title} {\bibinfo {title} {{QCD matter in extreme
  environments}},\ }\href {https://doi.org/10.1088/0954-3899/39/1/013101}
  {\bibfield  {journal} {\bibinfo  {journal} {J. Phys. G}\ }\textbf {\bibinfo
  {volume} {39}},\ \bibinfo {pages} {013101} (\bibinfo {year} {2012})},\
  \Eprint {https://arxiv.org/abs/1108.2939} {arXiv:1108.2939 [hep-ph]}
  \BibitemShut {NoStop}%
\bibitem [{\citenamefont {Kharzeev}\ \emph {et~al.}(2013)\citenamefont
  {Kharzeev}, \citenamefont {Landsteiner}, \citenamefont {Schmitt},\ and\
  \citenamefont {Yee}}]{Kharzeev:2013jha}%
  \BibitemOpen
  \bibinfo {editor} {\bibfnamefont {D.}~\bibnamefont {Kharzeev}}, \bibinfo
  {editor} {\bibfnamefont {K.}~\bibnamefont {Landsteiner}}, \bibinfo {editor}
  {\bibfnamefont {A.}~\bibnamefont {Schmitt}},\ and\ \bibinfo {editor}
  {\bibfnamefont {H.-U.}\ \bibnamefont {Yee}},\ eds.,\ \href
  {https://doi.org/10.1007/978-3-642-37305-3} {\emph {\bibinfo {title}
  {{Strongly Interacting Matter in Magnetic Fields}}}},\ Vol.\ \bibinfo
  {volume} {871}\ (\bibinfo {year} {2013})\BibitemShut {NoStop}%
\bibitem [{\citenamefont {Chen}\ \emph {et~al.}(2021)\citenamefont {Chen},
  \citenamefont {Huang},\ and\ \citenamefont {Liao}}]{Chen:2021aiq}%
  \BibitemOpen
  \bibfield  {author} {\bibinfo {author} {\bibfnamefont {H.-L.}\ \bibnamefont
  {Chen}}, \bibinfo {author} {\bibfnamefont {X.-G.}\ \bibnamefont {Huang}},\
  and\ \bibinfo {author} {\bibfnamefont {J.}~\bibnamefont {Liao}},\ }\bibfield
  {title} {\bibinfo {title} {{QCD phase structure under rotation}},\ }\href
  {https://doi.org/10.1007/978-3-030-71427-7_11} {\bibfield  {journal}
  {\bibinfo  {journal} {Lect. Notes Phys.}\ }\textbf {\bibinfo {volume}
  {987}},\ \bibinfo {pages} {349} (\bibinfo {year} {2021})},\ \Eprint
  {https://arxiv.org/abs/2108.00586} {arXiv:2108.00586 [hep-ph]} \BibitemShut
  {NoStop}%
\bibitem [{\citenamefont {Kharzeev}\ \emph {et~al.}(2008)\citenamefont
  {Kharzeev}, \citenamefont {McLerran},\ and\ \citenamefont
  {Warringa}}]{Kharzeev:2007jp}%
  \BibitemOpen
  \bibfield  {author} {\bibinfo {author} {\bibfnamefont {D.~E.}\ \bibnamefont
  {Kharzeev}}, \bibinfo {author} {\bibfnamefont {L.~D.}\ \bibnamefont
  {McLerran}},\ and\ \bibinfo {author} {\bibfnamefont {H.~J.}\ \bibnamefont
  {Warringa}},\ }\bibfield  {title} {\bibinfo {title} {{The Effects of
  topological charge change in heavy ion collisions: 'Event by event P and CP
  violation'}},\ }\href {https://doi.org/10.1016/j.nuclphysa.2008.02.298}
  {\bibfield  {journal} {\bibinfo  {journal} {Nucl. Phys. A}\ }\textbf
  {\bibinfo {volume} {803}},\ \bibinfo {pages} {227} (\bibinfo {year}
  {2008})},\ \Eprint {https://arxiv.org/abs/0711.0950} {arXiv:0711.0950
  [hep-ph]} \BibitemShut {NoStop}%
\bibitem [{\citenamefont {Skokov}\ \emph {et~al.}(2009)\citenamefont {Skokov},
  \citenamefont {Illarionov},\ and\ \citenamefont {Toneev}}]{Skokov:2009qp}%
  \BibitemOpen
  \bibfield  {author} {\bibinfo {author} {\bibfnamefont {V.}~\bibnamefont
  {Skokov}}, \bibinfo {author} {\bibfnamefont {A.~Y.}\ \bibnamefont
  {Illarionov}},\ and\ \bibinfo {author} {\bibfnamefont {V.}~\bibnamefont
  {Toneev}},\ }\bibfield  {title} {\bibinfo {title} {{Estimate of the magnetic
  field strength in heavy-ion collisions}},\ }\href
  {https://doi.org/10.1142/S0217751X09047570} {\bibfield  {journal} {\bibinfo
  {journal} {Int. J. Mod. Phys. A}\ }\textbf {\bibinfo {volume} {24}},\
  \bibinfo {pages} {5925} (\bibinfo {year} {2009})},\ \Eprint
  {https://arxiv.org/abs/0907.1396} {arXiv:0907.1396 [nucl-th]} \BibitemShut
  {NoStop}%
\bibitem [{\citenamefont {Becattini}\ \emph {et~al.}(2008)\citenamefont
  {Becattini}, \citenamefont {Piccinini},\ and\ \citenamefont
  {Rizzo}}]{Becattini:2007sr}%
  \BibitemOpen
  \bibfield  {author} {\bibinfo {author} {\bibfnamefont {F.}~\bibnamefont
  {Becattini}}, \bibinfo {author} {\bibfnamefont {F.}~\bibnamefont
  {Piccinini}},\ and\ \bibinfo {author} {\bibfnamefont {J.}~\bibnamefont
  {Rizzo}},\ }\bibfield  {title} {\bibinfo {title} {{Angular momentum
  conservation in heavy ion collisions at very high energy}},\ }\href
  {https://doi.org/10.1103/PhysRevC.77.024906} {\bibfield  {journal} {\bibinfo
  {journal} {Phys. Rev. C}\ }\textbf {\bibinfo {volume} {77}},\ \bibinfo
  {pages} {024906} (\bibinfo {year} {2008})},\ \Eprint
  {https://arxiv.org/abs/0711.1253} {arXiv:0711.1253 [nucl-th]} \BibitemShut
  {NoStop}%
\bibitem [{\citenamefont {Jiang}\ \emph {et~al.}(2016)\citenamefont {Jiang},
  \citenamefont {Lin},\ and\ \citenamefont {Liao}}]{Jiang:2016woz}%
  \BibitemOpen
  \bibfield  {author} {\bibinfo {author} {\bibfnamefont {Y.}~\bibnamefont
  {Jiang}}, \bibinfo {author} {\bibfnamefont {Z.-W.}\ \bibnamefont {Lin}},\
  and\ \bibinfo {author} {\bibfnamefont {J.}~\bibnamefont {Liao}},\ }\bibfield
  {title} {\bibinfo {title} {{Rotating quark-gluon plasma in relativistic heavy
  ion collisions}},\ }\href {https://doi.org/10.1103/PhysRevC.94.044910}
  {\bibfield  {journal} {\bibinfo  {journal} {Phys. Rev. C}\ }\textbf {\bibinfo
  {volume} {94}},\ \bibinfo {pages} {044910} (\bibinfo {year} {2016})},\
  \bibinfo {note} {[Erratum: Phys.Rev.C 95, 049904 (2017)]},\ \Eprint
  {https://arxiv.org/abs/1602.06580} {arXiv:1602.06580 [hep-ph]} \BibitemShut
  {NoStop}%
\bibitem [{\citenamefont {Kharzeev}\ and\ \citenamefont
  {Zhitnitsky}(2007)}]{Kharzeev:2007tn}%
  \BibitemOpen
  \bibfield  {author} {\bibinfo {author} {\bibfnamefont {D.}~\bibnamefont
  {Kharzeev}}\ and\ \bibinfo {author} {\bibfnamefont {A.}~\bibnamefont
  {Zhitnitsky}},\ }\bibfield  {title} {\bibinfo {title} {{Charge separation
  induced by P-odd bubbles in QCD matter}},\ }\href
  {https://doi.org/10.1016/j.nuclphysa.2007.10.001} {\bibfield  {journal}
  {\bibinfo  {journal} {Nucl. Phys. A}\ }\textbf {\bibinfo {volume} {797}},\
  \bibinfo {pages} {67} (\bibinfo {year} {2007})},\ \Eprint
  {https://arxiv.org/abs/0706.1026} {arXiv:0706.1026 [hep-ph]} \BibitemShut
  {NoStop}%
\bibitem [{\citenamefont {Fukushima}\ \emph {et~al.}(2008)\citenamefont
  {Fukushima}, \citenamefont {Kharzeev},\ and\ \citenamefont
  {Warringa}}]{Fukushima:2008xe}%
  \BibitemOpen
  \bibfield  {author} {\bibinfo {author} {\bibfnamefont {K.}~\bibnamefont
  {Fukushima}}, \bibinfo {author} {\bibfnamefont {D.~E.}\ \bibnamefont
  {Kharzeev}},\ and\ \bibinfo {author} {\bibfnamefont {H.~J.}\ \bibnamefont
  {Warringa}},\ }\bibfield  {title} {\bibinfo {title} {{The Chiral Magnetic
  Effect}},\ }\href {https://doi.org/10.1103/PhysRevD.78.074033} {\bibfield
  {journal} {\bibinfo  {journal} {Phys. Rev. D}\ }\textbf {\bibinfo {volume}
  {78}},\ \bibinfo {pages} {074033} (\bibinfo {year} {2008})},\ \Eprint
  {https://arxiv.org/abs/0808.3382} {arXiv:0808.3382 [hep-ph]} \BibitemShut
  {NoStop}%
\bibitem [{\citenamefont {Kharzeev}\ and\ \citenamefont
  {Son}(2011)}]{Kharzeev:2010gr}%
  \BibitemOpen
  \bibfield  {author} {\bibinfo {author} {\bibfnamefont {D.~E.}\ \bibnamefont
  {Kharzeev}}\ and\ \bibinfo {author} {\bibfnamefont {D.~T.}\ \bibnamefont
  {Son}},\ }\bibfield  {title} {\bibinfo {title} {{Testing the chiral magnetic
  and chiral vortical effects in heavy ion collisions}},\ }\href
  {https://doi.org/10.1103/PhysRevLett.106.062301} {\bibfield  {journal}
  {\bibinfo  {journal} {Phys. Rev. Lett.}\ }\textbf {\bibinfo {volume} {106}},\
  \bibinfo {pages} {062301} (\bibinfo {year} {2011})},\ \Eprint
  {https://arxiv.org/abs/1010.0038} {arXiv:1010.0038 [hep-ph]} \BibitemShut
  {NoStop}%
\bibitem [{\citenamefont {Klevansky}\ and\ \citenamefont
  {Lemmer}(1989)}]{Klevansky:1989vi}%
  \BibitemOpen
  \bibfield  {author} {\bibinfo {author} {\bibfnamefont {S.~P.}\ \bibnamefont
  {Klevansky}}\ and\ \bibinfo {author} {\bibfnamefont {R.~H.}\ \bibnamefont
  {Lemmer}},\ }\bibfield  {title} {\bibinfo {title} {{Chiral symmetry
  restoration in the Nambu-Jona-Lasinio model with a constant electromagnetic
  field}},\ }\href {https://doi.org/10.1103/PhysRevD.39.3478} {\bibfield
  {journal} {\bibinfo  {journal} {Phys. Rev. D}\ }\textbf {\bibinfo {volume}
  {39}},\ \bibinfo {pages} {3478} (\bibinfo {year} {1989})}\BibitemShut
  {NoStop}%
\bibitem [{\citenamefont {Klimenko}(1991)}]{Klimenko:1990rh}%
  \BibitemOpen
  \bibfield  {author} {\bibinfo {author} {\bibfnamefont {K.~G.}\ \bibnamefont
  {Klimenko}},\ }\bibfield  {title} {\bibinfo {title} {{Three-dimensional
  Gross-Neveu model in an external magnetic field}},\ }\href
  {https://doi.org/10.1007/BF01015908} {\bibfield  {journal} {\bibinfo
  {journal} {Teor. Mat. Fiz.}\ }\textbf {\bibinfo {volume} {89}},\ \bibinfo
  {pages} {211} (\bibinfo {year} {1991})}\BibitemShut {NoStop}%
\bibitem [{\citenamefont {Gusynin}\ \emph {et~al.}(1996)\citenamefont
  {Gusynin}, \citenamefont {Miransky},\ and\ \citenamefont
  {Shovkovy}}]{Gusynin:1995nb}%
  \BibitemOpen
  \bibfield  {author} {\bibinfo {author} {\bibfnamefont {V.~P.}\ \bibnamefont
  {Gusynin}}, \bibinfo {author} {\bibfnamefont {V.~A.}\ \bibnamefont
  {Miransky}},\ and\ \bibinfo {author} {\bibfnamefont {I.~A.}\ \bibnamefont
  {Shovkovy}},\ }\bibfield  {title} {\bibinfo {title} {{Dimensional reduction
  and catalysis of dynamical symmetry breaking by a magnetic field}},\ }\href
  {https://doi.org/10.1016/0550-3213(96)00021-1} {\bibfield  {journal}
  {\bibinfo  {journal} {Nucl. Phys. B}\ }\textbf {\bibinfo {volume} {462}},\
  \bibinfo {pages} {249} (\bibinfo {year} {1996})},\ \Eprint
  {https://arxiv.org/abs/hep-ph/9509320} {arXiv:hep-ph/9509320} \BibitemShut
  {NoStop}%
\bibitem [{\citenamefont {Bali}\ \emph
  {et~al.}(2012{\natexlab{a}})\citenamefont {Bali}, \citenamefont {Bruckmann},
  \citenamefont {Endrodi}, \citenamefont {Fodor}, \citenamefont {Katz},
  \citenamefont {Krieg}, \citenamefont {Schafer},\ and\ \citenamefont
  {Szabo}}]{Bali:2011qj}%
  \BibitemOpen
  \bibfield  {author} {\bibinfo {author} {\bibfnamefont {G.~S.}\ \bibnamefont
  {Bali}}, \bibinfo {author} {\bibfnamefont {F.}~\bibnamefont {Bruckmann}},
  \bibinfo {author} {\bibfnamefont {G.}~\bibnamefont {Endrodi}}, \bibinfo
  {author} {\bibfnamefont {Z.}~\bibnamefont {Fodor}}, \bibinfo {author}
  {\bibfnamefont {S.~D.}\ \bibnamefont {Katz}}, \bibinfo {author}
  {\bibfnamefont {S.}~\bibnamefont {Krieg}}, \bibinfo {author} {\bibfnamefont
  {A.}~\bibnamefont {Schafer}},\ and\ \bibinfo {author} {\bibfnamefont {K.~K.}\
  \bibnamefont {Szabo}},\ }\bibfield  {title} {\bibinfo {title} {{The QCD phase
  diagram for external magnetic fields}},\ }\href
  {https://doi.org/10.1007/JHEP02(2012)044} {\bibfield  {journal} {\bibinfo
  {journal} {JHEP}\ }\textbf {\bibinfo {volume} {02}},\ \bibinfo {pages}
  {044}},\ \Eprint {https://arxiv.org/abs/1111.4956} {arXiv:1111.4956
  [hep-lat]} \BibitemShut {NoStop}%
\bibitem [{\citenamefont {Bali}\ \emph
  {et~al.}(2012{\natexlab{b}})\citenamefont {Bali}, \citenamefont {Bruckmann},
  \citenamefont {Endrodi}, \citenamefont {Fodor}, \citenamefont {Katz},\ and\
  \citenamefont {Schafer}}]{Bali:2012zg}%
  \BibitemOpen
  \bibfield  {author} {\bibinfo {author} {\bibfnamefont {G.~S.}\ \bibnamefont
  {Bali}}, \bibinfo {author} {\bibfnamefont {F.}~\bibnamefont {Bruckmann}},
  \bibinfo {author} {\bibfnamefont {G.}~\bibnamefont {Endrodi}}, \bibinfo
  {author} {\bibfnamefont {Z.}~\bibnamefont {Fodor}}, \bibinfo {author}
  {\bibfnamefont {S.~D.}\ \bibnamefont {Katz}},\ and\ \bibinfo {author}
  {\bibfnamefont {A.}~\bibnamefont {Schafer}},\ }\bibfield  {title} {\bibinfo
  {title} {{QCD quark condensate in external magnetic fields}},\ }\href
  {https://doi.org/10.1103/PhysRevD.86.071502} {\bibfield  {journal} {\bibinfo
  {journal} {Phys. Rev. D}\ }\textbf {\bibinfo {volume} {86}},\ \bibinfo
  {pages} {071502} (\bibinfo {year} {2012}{\natexlab{b}})},\ \Eprint
  {https://arxiv.org/abs/1206.4205} {arXiv:1206.4205 [hep-lat]} \BibitemShut
  {NoStop}%
\bibitem [{\citenamefont {Bali}\ \emph {et~al.}(2013)\citenamefont {Bali},
  \citenamefont {Bruckmann}, \citenamefont {Endrodi}, \citenamefont {Gruber},\
  and\ \citenamefont {Schaefer}}]{Bali:2013esa}%
  \BibitemOpen
  \bibfield  {author} {\bibinfo {author} {\bibfnamefont {G.~S.}\ \bibnamefont
  {Bali}}, \bibinfo {author} {\bibfnamefont {F.}~\bibnamefont {Bruckmann}},
  \bibinfo {author} {\bibfnamefont {G.}~\bibnamefont {Endrodi}}, \bibinfo
  {author} {\bibfnamefont {F.}~\bibnamefont {Gruber}},\ and\ \bibinfo {author}
  {\bibfnamefont {A.}~\bibnamefont {Schaefer}},\ }\bibfield  {title} {\bibinfo
  {title} {{Magnetic field-induced gluonic (inverse) catalysis and pressure
  (an)isotropy in QCD}},\ }\href {https://doi.org/10.1007/JHEP04(2013)130}
  {\bibfield  {journal} {\bibinfo  {journal} {JHEP}\ }\textbf {\bibinfo
  {volume} {04}},\ \bibinfo {pages} {130}},\ \Eprint
  {https://arxiv.org/abs/1303.1328} {arXiv:1303.1328 [hep-lat]} \BibitemShut
  {NoStop}%
\bibitem [{\citenamefont {Miransky}\ and\ \citenamefont
  {Shovkovy}(2015)}]{Miransky:2015ava}%
  \BibitemOpen
  \bibfield  {author} {\bibinfo {author} {\bibfnamefont {V.~A.}\ \bibnamefont
  {Miransky}}\ and\ \bibinfo {author} {\bibfnamefont {I.~A.}\ \bibnamefont
  {Shovkovy}},\ }\bibfield  {title} {\bibinfo {title} {{Quantum field theory in
  a magnetic field: From quantum chromodynamics to graphene and Dirac
  semimetals}},\ }\href {https://doi.org/10.1016/j.physrep.2015.02.003}
  {\bibfield  {journal} {\bibinfo  {journal} {Phys. Rept.}\ }\textbf {\bibinfo
  {volume} {576}},\ \bibinfo {pages} {1} (\bibinfo {year} {2015})},\ \Eprint
  {https://arxiv.org/abs/1503.00732} {arXiv:1503.00732 [hep-ph]} \BibitemShut
  {NoStop}%
\bibitem [{\citenamefont {Son}\ and\ \citenamefont
  {Surowka}(2009)}]{Son:2009tf}%
  \BibitemOpen
  \bibfield  {author} {\bibinfo {author} {\bibfnamefont {D.~T.}\ \bibnamefont
  {Son}}\ and\ \bibinfo {author} {\bibfnamefont {P.}~\bibnamefont {Surowka}},\
  }\bibfield  {title} {\bibinfo {title} {{Hydrodynamics with Triangle
  Anomalies}},\ }\href {https://doi.org/10.1103/PhysRevLett.103.191601}
  {\bibfield  {journal} {\bibinfo  {journal} {Phys. Rev. Lett.}\ }\textbf
  {\bibinfo {volume} {103}},\ \bibinfo {pages} {191601} (\bibinfo {year}
  {2009})},\ \Eprint {https://arxiv.org/abs/0906.5044} {arXiv:0906.5044
  [hep-th]} \BibitemShut {NoStop}%
\bibitem [{\citenamefont {Jiang}\ \emph {et~al.}(2015)\citenamefont {Jiang},
  \citenamefont {Huang},\ and\ \citenamefont {Liao}}]{Jiang:2015cva}%
  \BibitemOpen
  \bibfield  {author} {\bibinfo {author} {\bibfnamefont {Y.}~\bibnamefont
  {Jiang}}, \bibinfo {author} {\bibfnamefont {X.-G.}\ \bibnamefont {Huang}},\
  and\ \bibinfo {author} {\bibfnamefont {J.}~\bibnamefont {Liao}},\ }\bibfield
  {title} {\bibinfo {title} {{Chiral vortical wave and induced flavor charge
  transport in a rotating quark-gluon plasma}},\ }\href
  {https://doi.org/10.1103/PhysRevD.92.071501} {\bibfield  {journal} {\bibinfo
  {journal} {Phys. Rev. D}\ }\textbf {\bibinfo {volume} {92}},\ \bibinfo
  {pages} {071501} (\bibinfo {year} {2015})},\ \Eprint
  {https://arxiv.org/abs/1504.03201} {arXiv:1504.03201 [hep-ph]} \BibitemShut
  {NoStop}%
\bibitem [{\citenamefont {Kharzeev}\ \emph {et~al.}(2016)\citenamefont
  {Kharzeev}, \citenamefont {Liao}, \citenamefont {Voloshin},\ and\
  \citenamefont {Wang}}]{Kharzeev:2015znc}%
  \BibitemOpen
  \bibfield  {author} {\bibinfo {author} {\bibfnamefont {D.~E.}\ \bibnamefont
  {Kharzeev}}, \bibinfo {author} {\bibfnamefont {J.}~\bibnamefont {Liao}},
  \bibinfo {author} {\bibfnamefont {S.~A.}\ \bibnamefont {Voloshin}},\ and\
  \bibinfo {author} {\bibfnamefont {G.}~\bibnamefont {Wang}},\ }\bibfield
  {title} {\bibinfo {title} {{Chiral magnetic and vortical effects in
  high-energy nuclear collisions\textemdash{}A status report}},\ }\href
  {https://doi.org/10.1016/j.ppnp.2016.01.001} {\bibfield  {journal} {\bibinfo
  {journal} {Prog. Part. Nucl. Phys.}\ }\textbf {\bibinfo {volume} {88}},\
  \bibinfo {pages} {1} (\bibinfo {year} {2016})},\ \Eprint
  {https://arxiv.org/abs/1511.04050} {arXiv:1511.04050 [hep-ph]} \BibitemShut
  {NoStop}%
\bibitem [{\citenamefont {Jiang}\ and\ \citenamefont
  {Liao}(2016)}]{Jiang:2016wvv}%
  \BibitemOpen
  \bibfield  {author} {\bibinfo {author} {\bibfnamefont {Y.}~\bibnamefont
  {Jiang}}\ and\ \bibinfo {author} {\bibfnamefont {J.}~\bibnamefont {Liao}},\
  }\bibfield  {title} {\bibinfo {title} {{Pairing Phase Transitions of Matter
  under Rotation}},\ }\href {https://doi.org/10.1103/PhysRevLett.117.192302}
  {\bibfield  {journal} {\bibinfo  {journal} {Phys. Rev. Lett.}\ }\textbf
  {\bibinfo {volume} {117}},\ \bibinfo {pages} {192302} (\bibinfo {year}
  {2016})},\ \Eprint {https://arxiv.org/abs/1606.03808} {arXiv:1606.03808
  [hep-ph]} \BibitemShut {NoStop}%
\bibitem [{\citenamefont {Chernodub}\ and\ \citenamefont
  {Gongyo}(2017)}]{Chernodub:2016kxh}%
  \BibitemOpen
  \bibfield  {author} {\bibinfo {author} {\bibfnamefont {M.~N.}\ \bibnamefont
  {Chernodub}}\ and\ \bibinfo {author} {\bibfnamefont {S.}~\bibnamefont
  {Gongyo}},\ }\bibfield  {title} {\bibinfo {title} {{Interacting fermions in
  rotation: chiral symmetry restoration, moment of inertia and
  thermodynamics}},\ }\href {https://doi.org/10.1007/JHEP01(2017)136}
  {\bibfield  {journal} {\bibinfo  {journal} {JHEP}\ }\textbf {\bibinfo
  {volume} {01}},\ \bibinfo {pages} {136}},\ \Eprint
  {https://arxiv.org/abs/1611.02598} {arXiv:1611.02598 [hep-th]} \BibitemShut
  {NoStop}%
\bibitem [{\citenamefont {Wang}\ \emph {et~al.}(2019)\citenamefont {Wang},
  \citenamefont {Wei}, \citenamefont {Li},\ and\ \citenamefont
  {Huang}}]{Wang:2018sur}%
  \BibitemOpen
  \bibfield  {author} {\bibinfo {author} {\bibfnamefont {X.}~\bibnamefont
  {Wang}}, \bibinfo {author} {\bibfnamefont {M.}~\bibnamefont {Wei}}, \bibinfo
  {author} {\bibfnamefont {Z.}~\bibnamefont {Li}},\ and\ \bibinfo {author}
  {\bibfnamefont {M.}~\bibnamefont {Huang}},\ }\bibfield  {title} {\bibinfo
  {title} {{Quark matter under rotation in the NJL model with vector
  interaction}},\ }\href {https://doi.org/10.1103/PhysRevD.99.016018}
  {\bibfield  {journal} {\bibinfo  {journal} {Phys. Rev. D}\ }\textbf {\bibinfo
  {volume} {99}},\ \bibinfo {pages} {016018} (\bibinfo {year} {2019})},\
  \Eprint {https://arxiv.org/abs/1808.01931} {arXiv:1808.01931 [hep-ph]}
  \BibitemShut {NoStop}%
\bibitem [{\citenamefont {Jiang}(2021)}]{Jiang:2021izj}%
  \BibitemOpen
  \bibfield  {author} {\bibinfo {author} {\bibfnamefont {Y.}~\bibnamefont
  {Jiang}},\ }\bibfield  {title} {\bibinfo {title} {{Chiral vortical
  catalysis}},\ }\href@noop {} {\  (\bibinfo {year} {2021})},\ \Eprint
  {https://arxiv.org/abs/2108.09622} {arXiv:2108.09622 [hep-ph]} \BibitemShut
  {NoStop}%
\bibitem [{\citenamefont {Chernodub}(2021)}]{Chernodub:2020qah}%
  \BibitemOpen
  \bibfield  {author} {\bibinfo {author} {\bibfnamefont {M.~N.}\ \bibnamefont
  {Chernodub}},\ }\bibfield  {title} {\bibinfo {title} {{Inhomogeneous
  confining-deconfining phases in rotating plasmas}},\ }\href
  {https://doi.org/10.1103/PhysRevD.103.054027} {\bibfield  {journal} {\bibinfo
   {journal} {Phys. Rev. D}\ }\textbf {\bibinfo {volume} {103}},\ \bibinfo
  {pages} {054027} (\bibinfo {year} {2021})},\ \Eprint
  {https://arxiv.org/abs/2012.04924} {arXiv:2012.04924 [hep-ph]} \BibitemShut
  {NoStop}%
\bibitem [{\citenamefont {Braguta}\ \emph {et~al.}(2021)\citenamefont
  {Braguta}, \citenamefont {Kotov}, \citenamefont {Kuznedelev},\ and\
  \citenamefont {Roenko}}]{Braguta:2021jgn}%
  \BibitemOpen
  \bibfield  {author} {\bibinfo {author} {\bibfnamefont {V.~V.}\ \bibnamefont
  {Braguta}}, \bibinfo {author} {\bibfnamefont {A.~Y.}\ \bibnamefont {Kotov}},
  \bibinfo {author} {\bibfnamefont {D.~D.}\ \bibnamefont {Kuznedelev}},\ and\
  \bibinfo {author} {\bibfnamefont {A.~A.}\ \bibnamefont {Roenko}},\ }\bibfield
   {title} {\bibinfo {title} {{Influence of relativistic rotation on the
  confinement-deconfinement transition in gluodynamics}},\ }\href
  {https://doi.org/10.1103/PhysRevD.103.094515} {\bibfield  {journal} {\bibinfo
   {journal} {Phys. Rev. D}\ }\textbf {\bibinfo {volume} {103}},\ \bibinfo
  {pages} {094515} (\bibinfo {year} {2021})},\ \Eprint
  {https://arxiv.org/abs/2102.05084} {arXiv:2102.05084 [hep-lat]} \BibitemShut
  {NoStop}%
\bibitem [{\citenamefont {Fetter}(2009)}]{Fetter:2009zz}%
  \BibitemOpen
  \bibfield  {author} {\bibinfo {author} {\bibfnamefont {A.~L.}\ \bibnamefont
  {Fetter}},\ }\bibfield  {title} {\bibinfo {title} {{Rotating trapped
  Bose-Einstein condensates}},\ }\href
  {https://doi.org/10.1103/RevModPhys.81.647} {\bibfield  {journal} {\bibinfo
  {journal} {Rev. Mod. Phys.}\ }\textbf {\bibinfo {volume} {81}},\ \bibinfo
  {pages} {647} (\bibinfo {year} {2009})}\BibitemShut {NoStop}%
\bibitem [{\citenamefont {Berti}\ \emph {et~al.}(2005)\citenamefont {Berti},
  \citenamefont {White}, \citenamefont {Maniopoulou},\ and\ \citenamefont
  {Bruni}}]{Berti:2004ny}%
  \BibitemOpen
  \bibfield  {author} {\bibinfo {author} {\bibfnamefont {E.}~\bibnamefont
  {Berti}}, \bibinfo {author} {\bibfnamefont {F.}~\bibnamefont {White}},
  \bibinfo {author} {\bibfnamefont {A.}~\bibnamefont {Maniopoulou}},\ and\
  \bibinfo {author} {\bibfnamefont {M.}~\bibnamefont {Bruni}},\ }\bibfield
  {title} {\bibinfo {title} {{Rotating neutron stars: An Invariant comparison
  of approximate and numerical spacetime models}},\ }\href
  {https://doi.org/10.1111/j.1365-2966.2005.08812.x} {\bibfield  {journal}
  {\bibinfo  {journal} {Mon. Not. Roy. Astron. Soc.}\ }\textbf {\bibinfo
  {volume} {358}},\ \bibinfo {pages} {923} (\bibinfo {year} {2005})},\ \Eprint
  {https://arxiv.org/abs/gr-qc/0405146} {arXiv:gr-qc/0405146} \BibitemShut
  {NoStop}%
\bibitem [{\citenamefont {Adamczyk}\ \emph {et~al.}(2017)\citenamefont
  {Adamczyk} \emph {et~al.}}]{STAR:2017ckg}%
  \BibitemOpen
  \bibfield  {author} {\bibinfo {author} {\bibfnamefont {L.}~\bibnamefont
  {Adamczyk}} \emph {et~al.} (\bibinfo {collaboration} {STAR}),\ }\bibfield
  {title} {\bibinfo {title} {{Global $\Lambda$ hyperon polarization in nuclear
  collisions: evidence for the most vortical fluid}},\ }\href
  {https://doi.org/10.1038/nature23004} {\bibfield  {journal} {\bibinfo
  {journal} {Nature}\ }\textbf {\bibinfo {volume} {548}},\ \bibinfo {pages}
  {62} (\bibinfo {year} {2017})},\ \Eprint {https://arxiv.org/abs/1701.06657}
  {arXiv:1701.06657 [nucl-ex]} \BibitemShut {NoStop}%
\bibitem [{\citenamefont {Acharya}\ \emph {et~al.}(2020)\citenamefont {Acharya}
  \emph {et~al.}}]{ALICE:2019aid}%
  \BibitemOpen
  \bibfield  {author} {\bibinfo {author} {\bibfnamefont {S.}~\bibnamefont
  {Acharya}} \emph {et~al.} (\bibinfo {collaboration} {ALICE}),\ }\bibfield
  {title} {\bibinfo {title} {{Evidence of Spin-Orbital Angular Momentum
  Interactions in Relativistic Heavy-Ion Collisions}},\ }\href
  {https://doi.org/10.1103/PhysRevLett.125.012301} {\bibfield  {journal}
  {\bibinfo  {journal} {Phys. Rev. Lett.}\ }\textbf {\bibinfo {volume} {125}},\
  \bibinfo {pages} {012301} (\bibinfo {year} {2020})},\ \Eprint
  {https://arxiv.org/abs/1910.14408} {arXiv:1910.14408 [nucl-ex]} \BibitemShut
  {NoStop}%
\bibitem [{\citenamefont {Adare}\ \emph {et~al.}(2012)\citenamefont {Adare}
  \emph {et~al.}}]{PHENIX:2011oxq}%
  \BibitemOpen
  \bibfield  {author} {\bibinfo {author} {\bibfnamefont {A.}~\bibnamefont
  {Adare}} \emph {et~al.} (\bibinfo {collaboration} {PHENIX}),\ }\bibfield
  {title} {\bibinfo {title} {{Observation of direct-photon collective flow in
  $\sqrt{s_{NN}}=200$ GeV Au+Au collisions}},\ }\href
  {https://doi.org/10.1103/PhysRevLett.109.122302} {\bibfield  {journal}
  {\bibinfo  {journal} {Phys. Rev. Lett.}\ }\textbf {\bibinfo {volume} {109}},\
  \bibinfo {pages} {122302} (\bibinfo {year} {2012})},\ \Eprint
  {https://arxiv.org/abs/1105.4126} {arXiv:1105.4126 [nucl-ex]} \BibitemShut
  {NoStop}%
\bibitem [{\citenamefont {Lohner}(2013)}]{Lohner:2012ct}%
  \BibitemOpen
  \bibfield  {author} {\bibinfo {author} {\bibfnamefont {D.}~\bibnamefont
  {Lohner}} (\bibinfo {collaboration} {ALICE}),\ }\bibfield  {title} {\bibinfo
  {title} {{Measurement of Direct-Photon Elliptic Flow in Pb-Pb Collisions at
  $\sqrt{s_{NN}} = 2.76$ TeV}},\ }\href
  {https://doi.org/10.1088/1742-6596/446/1/012028} {\bibfield  {journal}
  {\bibinfo  {journal} {J. Phys. Conf. Ser.}\ }\textbf {\bibinfo {volume}
  {446}},\ \bibinfo {pages} {012028} (\bibinfo {year} {2013})},\ \Eprint
  {https://arxiv.org/abs/1212.3995} {arXiv:1212.3995 [hep-ex]} \BibitemShut
  {NoStop}%
\bibitem [{\citenamefont {Wang}\ \emph {et~al.}(2020)\citenamefont {Wang},
  \citenamefont {Shovkovy}, \citenamefont {Yu},\ and\ \citenamefont
  {Huang}}]{Wang:2020dsr}%
  \BibitemOpen
  \bibfield  {author} {\bibinfo {author} {\bibfnamefont {X.}~\bibnamefont
  {Wang}}, \bibinfo {author} {\bibfnamefont {I.~A.}\ \bibnamefont {Shovkovy}},
  \bibinfo {author} {\bibfnamefont {L.}~\bibnamefont {Yu}},\ and\ \bibinfo
  {author} {\bibfnamefont {M.}~\bibnamefont {Huang}},\ }\bibfield  {title}
  {\bibinfo {title} {{Ellipticity of photon emission from strongly magnetized
  hot QCD plasma}},\ }\href {https://doi.org/10.1103/PhysRevD.102.076010}
  {\bibfield  {journal} {\bibinfo  {journal} {Phys. Rev. D}\ }\textbf {\bibinfo
  {volume} {102}},\ \bibinfo {pages} {076010} (\bibinfo {year} {2020})},\
  \Eprint {https://arxiv.org/abs/2006.16254} {arXiv:2006.16254 [hep-ph]}
  \BibitemShut {NoStop}%
\bibitem [{\citenamefont {Wang}\ and\ \citenamefont
  {Shovkovy}(2021)}]{Wang:2021ebh}%
  \BibitemOpen
  \bibfield  {author} {\bibinfo {author} {\bibfnamefont {X.}~\bibnamefont
  {Wang}}\ and\ \bibinfo {author} {\bibfnamefont {I.}~\bibnamefont
  {Shovkovy}},\ }\bibfield  {title} {\bibinfo {title} {{Photon polarization
  tensor in a magnetized plasma: absorptive part}},\ }\href@noop {} {\
  (\bibinfo {year} {2021})},\ \Eprint {https://arxiv.org/abs/2103.01967}
  {arXiv:2103.01967 [nucl-th]} \BibitemShut {NoStop}%
\bibitem [{\citenamefont {Salabura}\ and\ \citenamefont
  {Stroth}(2021)}]{Salabura:2020tou}%
  \BibitemOpen
  \bibfield  {author} {\bibinfo {author} {\bibfnamefont {P.}~\bibnamefont
  {Salabura}}\ and\ \bibinfo {author} {\bibfnamefont {J.}~\bibnamefont
  {Stroth}},\ }\bibfield  {title} {\bibinfo {title} {{Dilepton radiation from
  strongly interacting systems}},\ }\href
  {https://doi.org/10.1016/j.ppnp.2021.103869} {\bibfield  {journal} {\bibinfo
  {journal} {Prog. Part. Nucl. Phys.}\ }\textbf {\bibinfo {volume} {120}},\
  \bibinfo {pages} {103869} (\bibinfo {year} {2021})},\ \Eprint
  {https://arxiv.org/abs/2005.14589} {arXiv:2005.14589 [nucl-ex]} \BibitemShut
  {NoStop}%
\bibitem [{\citenamefont {Forster}(2018)}]{Forster:1975pm}%
  \BibitemOpen
  \bibfield  {author} {\bibinfo {author} {\bibfnamefont {D.}~\bibnamefont
  {Forster}},\ }\href {https://doi.org/https://doi.org/10.1201/9780429493683}
  {\emph {\bibinfo {title} {{Hydrodynamics Fluctuation, Broken Symmetry and
  Correlation Function}}}}\ (\bibinfo  {publisher} {CRC Press},\ \bibinfo
  {year} {2018})\BibitemShut {NoStop}%
\bibitem [{\citenamefont {Davidson}\ and\ \citenamefont
  {Ruiz~Arriola}(1995)}]{Davidson:1995fq}%
  \BibitemOpen
  \bibfield  {author} {\bibinfo {author} {\bibfnamefont {R.~M.}\ \bibnamefont
  {Davidson}}\ and\ \bibinfo {author} {\bibfnamefont {E.}~\bibnamefont
  {Ruiz~Arriola}},\ }\bibfield  {title} {\bibinfo {title} {{Mesonic correlation
  functions in the NJL model with vector mesons}},\ }\href
  {https://doi.org/10.1016/0370-2693(95)01119-B} {\bibfield  {journal}
  {\bibinfo  {journal} {Phys. Lett. B}\ }\textbf {\bibinfo {volume} {359}},\
  \bibinfo {pages} {273} (\bibinfo {year} {1995})}\BibitemShut {NoStop}%
\bibitem [{\citenamefont {Kapusta}\ and\ \citenamefont
  {Gale}(2011)}]{Kapusta:2006pm}%
  \BibitemOpen
  \bibfield  {author} {\bibinfo {author} {\bibfnamefont {J.~I.}\ \bibnamefont
  {Kapusta}}\ and\ \bibinfo {author} {\bibfnamefont {C.}~\bibnamefont {Gale}},\
  }\href {https://doi.org/10.1017/CBO9780511535130} {\emph {\bibinfo {title}
  {{Finite-temperature field theory: Principles and applications}}}},\
  Cambridge Monographs on Mathematical Physics\ (\bibinfo  {publisher}
  {Cambridge University Press},\ \bibinfo {year} {2011})\BibitemShut {NoStop}%
\bibitem [{\citenamefont {Bellac}(2011)}]{Bellac:2011kqa}%
  \BibitemOpen
  \bibfield  {author} {\bibinfo {author} {\bibfnamefont {M.~L.}\ \bibnamefont
  {Bellac}},\ }\href {https://doi.org/10.1017/CBO9780511721700} {\emph
  {\bibinfo {title} {{Thermal Field Theory}}}},\ Cambridge Monographs on
  Mathematical Physics\ (\bibinfo  {publisher} {Cambridge University Press},\
  \bibinfo {year} {2011})\BibitemShut {NoStop}%
\bibitem [{\citenamefont {Francis}\ and\ \citenamefont
  {Kaczmarek}(2012)}]{Francis:2011bt}%
  \BibitemOpen
  \bibfield  {author} {\bibinfo {author} {\bibfnamefont {A.}~\bibnamefont
  {Francis}}\ and\ \bibinfo {author} {\bibfnamefont {O.}~\bibnamefont
  {Kaczmarek}},\ }\bibfield  {title} {\bibinfo {title} {{On the temperature
  dependence of the electrical conductivity in hot quenched lattice QCD}},\
  }\href {https://doi.org/10.1016/j.ppnp.2011.12.020} {\bibfield  {journal}
  {\bibinfo  {journal} {Prog. Part. Nucl. Phys.}\ }\textbf {\bibinfo {volume}
  {67}},\ \bibinfo {pages} {212} (\bibinfo {year} {2012})},\ \Eprint
  {https://arxiv.org/abs/1112.4802} {arXiv:1112.4802 [hep-lat]} \BibitemShut
  {NoStop}%
\bibitem [{\citenamefont {Kunihiro}(1991)}]{Kunihiro:1991qu}%
  \BibitemOpen
  \bibfield  {author} {\bibinfo {author} {\bibfnamefont {T.}~\bibnamefont
  {Kunihiro}},\ }\bibfield  {title} {\bibinfo {title} {{Quark number
  susceptibility and fluctuations in the vector channel at high
  temperatures}},\ }\href {https://doi.org/10.1016/0370-2693(91)90107-2}
  {\bibfield  {journal} {\bibinfo  {journal} {Phys. Lett. B}\ }\textbf
  {\bibinfo {volume} {271}},\ \bibinfo {pages} {395} (\bibinfo {year}
  {1991})}\BibitemShut {NoStop}%
\bibitem [{\citenamefont {Islam}\ \emph {et~al.}(2015)\citenamefont {Islam},
  \citenamefont {Majumder}, \citenamefont {Haque},\ and\ \citenamefont
  {Mustafa}}]{Islam:2014sea}%
  \BibitemOpen
  \bibfield  {author} {\bibinfo {author} {\bibfnamefont {C.~A.}\ \bibnamefont
  {Islam}}, \bibinfo {author} {\bibfnamefont {S.}~\bibnamefont {Majumder}},
  \bibinfo {author} {\bibfnamefont {N.}~\bibnamefont {Haque}},\ and\ \bibinfo
  {author} {\bibfnamefont {M.~G.}\ \bibnamefont {Mustafa}},\ }\bibfield
  {title} {\bibinfo {title} {{Vector meson spectral function and dilepton
  production rate in a hot and dense medium within an effective QCD
  approach}},\ }\href {https://doi.org/10.1007/JHEP02(2015)011} {\bibfield
  {journal} {\bibinfo  {journal} {JHEP}\ }\textbf {\bibinfo {volume} {02}},\
  \bibinfo {pages} {011}},\ \Eprint {https://arxiv.org/abs/1411.6407}
  {arXiv:1411.6407 [hep-ph]} \BibitemShut {NoStop}%
\bibitem [{\citenamefont {Gale}\ \emph {et~al.}(2015)\citenamefont {Gale},
  \citenamefont {Hidaka}, \citenamefont {Jeon}, \citenamefont {Lin},
  \citenamefont {Paquet}, \citenamefont {Pisarski}, \citenamefont {Satow},
  \citenamefont {Skokov},\ and\ \citenamefont {Vujanovic}}]{Gale:2014dfa}%
  \BibitemOpen
  \bibfield  {author} {\bibinfo {author} {\bibfnamefont {C.}~\bibnamefont
  {Gale}}, \bibinfo {author} {\bibfnamefont {Y.}~\bibnamefont {Hidaka}},
  \bibinfo {author} {\bibfnamefont {S.}~\bibnamefont {Jeon}}, \bibinfo {author}
  {\bibfnamefont {S.}~\bibnamefont {Lin}}, \bibinfo {author} {\bibfnamefont
  {J.-F.}\ \bibnamefont {Paquet}}, \bibinfo {author} {\bibfnamefont {R.~D.}\
  \bibnamefont {Pisarski}}, \bibinfo {author} {\bibfnamefont {D.}~\bibnamefont
  {Satow}}, \bibinfo {author} {\bibfnamefont {V.~V.}\ \bibnamefont {Skokov}},\
  and\ \bibinfo {author} {\bibfnamefont {G.}~\bibnamefont {Vujanovic}},\
  }\bibfield  {title} {\bibinfo {title} {{Production and Elliptic Flow of
  Dileptons and Photons in a Matrix Model of the Quark-Gluon Plasma}},\ }\href
  {https://doi.org/10.1103/PhysRevLett.114.072301} {\bibfield  {journal}
  {\bibinfo  {journal} {Phys. Rev. Lett.}\ }\textbf {\bibinfo {volume} {114}},\
  \bibinfo {pages} {072301} (\bibinfo {year} {2015})},\ \Eprint
  {https://arxiv.org/abs/1409.4778} {arXiv:1409.4778 [hep-ph]} \BibitemShut
  {NoStop}%
\bibitem [{\citenamefont {Hidaka}\ \emph {et~al.}(2015)\citenamefont {Hidaka},
  \citenamefont {Lin}, \citenamefont {Pisarski},\ and\ \citenamefont
  {Satow}}]{Hidaka:2015ima}%
  \BibitemOpen
  \bibfield  {author} {\bibinfo {author} {\bibfnamefont {Y.}~\bibnamefont
  {Hidaka}}, \bibinfo {author} {\bibfnamefont {S.}~\bibnamefont {Lin}},
  \bibinfo {author} {\bibfnamefont {R.~D.}\ \bibnamefont {Pisarski}},\ and\
  \bibinfo {author} {\bibfnamefont {D.}~\bibnamefont {Satow}},\ }\bibfield
  {title} {\bibinfo {title} {{Dilepton and photon production in the presence of
  a nontrivial Polyakov loop}},\ }\href
  {https://doi.org/10.1007/JHEP10(2015)005} {\bibfield  {journal} {\bibinfo
  {journal} {JHEP}\ }\textbf {\bibinfo {volume} {10}},\ \bibinfo {pages}
  {005}},\ \Eprint {https://arxiv.org/abs/1504.01770} {arXiv:1504.01770
  [hep-ph]} \BibitemShut {NoStop}%
\bibitem [{\citenamefont {Sadooghi}\ and\ \citenamefont
  {Taghinavaz}(2017)}]{Sadooghi:2016jyf}%
  \BibitemOpen
  \bibfield  {author} {\bibinfo {author} {\bibfnamefont {N.}~\bibnamefont
  {Sadooghi}}\ and\ \bibinfo {author} {\bibfnamefont {F.}~\bibnamefont
  {Taghinavaz}},\ }\bibfield  {title} {\bibinfo {title} {{Dilepton production
  rate in a hot and magnetized quark-gluon plasma}},\ }\href
  {https://doi.org/10.1016/j.aop.2016.11.008} {\bibfield  {journal} {\bibinfo
  {journal} {Annals Phys.}\ }\textbf {\bibinfo {volume} {376}},\ \bibinfo
  {pages} {218} (\bibinfo {year} {2017})},\ \Eprint
  {https://arxiv.org/abs/1601.04887} {arXiv:1601.04887 [hep-ph]} \BibitemShut
  {NoStop}%
\bibitem [{\citenamefont {Bandyopadhyay}\ \emph {et~al.}(2016)\citenamefont
  {Bandyopadhyay}, \citenamefont {Islam},\ and\ \citenamefont
  {Mustafa}}]{Bandyopadhyay:2016fyd}%
  \BibitemOpen
  \bibfield  {author} {\bibinfo {author} {\bibfnamefont {A.}~\bibnamefont
  {Bandyopadhyay}}, \bibinfo {author} {\bibfnamefont {C.~A.}\ \bibnamefont
  {Islam}},\ and\ \bibinfo {author} {\bibfnamefont {M.~G.}\ \bibnamefont
  {Mustafa}},\ }\bibfield  {title} {\bibinfo {title} {{Electromagnetic spectral
  properties and Debye screening of a strongly magnetized hot medium}},\ }\href
  {https://doi.org/10.1103/PhysRevD.94.114034} {\bibfield  {journal} {\bibinfo
  {journal} {Phys. Rev. D}\ }\textbf {\bibinfo {volume} {94}},\ \bibinfo
  {pages} {114034} (\bibinfo {year} {2016})},\ \Eprint
  {https://arxiv.org/abs/1602.06769} {arXiv:1602.06769 [hep-ph]} \BibitemShut
  {NoStop}%
\bibitem [{\citenamefont {Das}\ \emph {et~al.}(2021)\citenamefont {Das},
  \citenamefont {Bandyopadhyay},\ and\ \citenamefont {Islam}}]{Das:2021fma}%
  \BibitemOpen
  \bibfield  {author} {\bibinfo {author} {\bibfnamefont {A.}~\bibnamefont
  {Das}}, \bibinfo {author} {\bibfnamefont {A.}~\bibnamefont {Bandyopadhyay}},\
  and\ \bibinfo {author} {\bibfnamefont {C.~A.}\ \bibnamefont {Islam}},\
  }\bibfield  {title} {\bibinfo {title} {{Lepton pair production from a hot and
  dense QCD medium in presence of an arbitrary magnetic field}},\ }\href@noop
  {} {\  (\bibinfo {year} {2021})},\ \Eprint {https://arxiv.org/abs/2109.00019}
  {arXiv:2109.00019 [hep-ph]} \BibitemShut {NoStop}%
\bibitem [{\citenamefont {Bandyopadhyay}\ and\ \citenamefont
  {Mallik}(2017)}]{Bandyopadhyay:2017raf}%
  \BibitemOpen
  \bibfield  {author} {\bibinfo {author} {\bibfnamefont {A.}~\bibnamefont
  {Bandyopadhyay}}\ and\ \bibinfo {author} {\bibfnamefont {S.}~\bibnamefont
  {Mallik}},\ }\bibfield  {title} {\bibinfo {title} {{Effect of magnetic field
  on dilepton production in a hot plasma}},\ }\href
  {https://doi.org/10.1103/PhysRevD.95.074019} {\bibfield  {journal} {\bibinfo
  {journal} {Phys. Rev. D}\ }\textbf {\bibinfo {volume} {95}},\ \bibinfo
  {pages} {074019} (\bibinfo {year} {2017})},\ \Eprint
  {https://arxiv.org/abs/1704.01364} {arXiv:1704.01364 [hep-ph]} \BibitemShut
  {NoStop}%
\bibitem [{\citenamefont {Ghosh}\ and\ \citenamefont
  {Chandra}(2018)}]{Ghosh:2018xhh}%
  \BibitemOpen
  \bibfield  {author} {\bibinfo {author} {\bibfnamefont {S.}~\bibnamefont
  {Ghosh}}\ and\ \bibinfo {author} {\bibfnamefont {V.}~\bibnamefont
  {Chandra}},\ }\bibfield  {title} {\bibinfo {title} {{Electromagnetic spectral
  function and dilepton rate in a hot magnetized QCD medium}},\ }\href
  {https://doi.org/10.1103/PhysRevD.98.076006} {\bibfield  {journal} {\bibinfo
  {journal} {Phys. Rev. D}\ }\textbf {\bibinfo {volume} {98}},\ \bibinfo
  {pages} {076006} (\bibinfo {year} {2018})},\ \Eprint
  {https://arxiv.org/abs/1808.05176} {arXiv:1808.05176 [hep-ph]} \BibitemShut
  {NoStop}%
\bibitem [{\citenamefont {Islam}\ \emph {et~al.}(2019)\citenamefont {Islam},
  \citenamefont {Bandyopadhyay}, \citenamefont {Roy},\ and\ \citenamefont
  {Sarkar}}]{Islam:2018sog}%
  \BibitemOpen
  \bibfield  {author} {\bibinfo {author} {\bibfnamefont {C.~A.}\ \bibnamefont
  {Islam}}, \bibinfo {author} {\bibfnamefont {A.}~\bibnamefont
  {Bandyopadhyay}}, \bibinfo {author} {\bibfnamefont {P.~K.}\ \bibnamefont
  {Roy}},\ and\ \bibinfo {author} {\bibfnamefont {S.}~\bibnamefont {Sarkar}},\
  }\bibfield  {title} {\bibinfo {title} {{Spectral function and dilepton rate
  from a strongly magnetized hot and dense medium in light of mean field
  models}},\ }\href {https://doi.org/10.1103/PhysRevD.99.094028} {\bibfield
  {journal} {\bibinfo  {journal} {Phys. Rev. D}\ }\textbf {\bibinfo {volume}
  {99}},\ \bibinfo {pages} {094028} (\bibinfo {year} {2019})},\ \Eprint
  {https://arxiv.org/abs/1812.10380} {arXiv:1812.10380 [hep-ph]} \BibitemShut
  {NoStop}%
\bibitem [{\citenamefont {Ghosh}\ \emph {et~al.}(2020)\citenamefont {Ghosh},
  \citenamefont {Chaudhuri}, \citenamefont {Sarkar},\ and\ \citenamefont
  {Roy}}]{Ghosh:2020xwp}%
  \BibitemOpen
  \bibfield  {author} {\bibinfo {author} {\bibfnamefont {S.}~\bibnamefont
  {Ghosh}}, \bibinfo {author} {\bibfnamefont {N.}~\bibnamefont {Chaudhuri}},
  \bibinfo {author} {\bibfnamefont {S.}~\bibnamefont {Sarkar}},\ and\ \bibinfo
  {author} {\bibfnamefont {P.}~\bibnamefont {Roy}},\ }\bibfield  {title}
  {\bibinfo {title} {{Effects of the anomalous magnetic moment of quarks on the
  dilepton production from hot and dense magnetized quark matter using the NJL
  model}},\ }\href {https://doi.org/10.1103/PhysRevD.101.096002} {\bibfield
  {journal} {\bibinfo  {journal} {Phys. Rev. D}\ }\textbf {\bibinfo {volume}
  {101}},\ \bibinfo {pages} {096002} (\bibinfo {year} {2020})},\ \Eprint
  {https://arxiv.org/abs/2004.09203} {arXiv:2004.09203 [nucl-th]} \BibitemShut
  {NoStop}%
\bibitem [{\citenamefont {Xu}\ \emph {et~al.}(2014)\citenamefont {Xu},
  \citenamefont {Pang},\ and\ \citenamefont {Wang}}]{Xu:2014ada}%
  \BibitemOpen
  \bibfield  {author} {\bibinfo {author} {\bibfnamefont {H.-j.}\ \bibnamefont
  {Xu}}, \bibinfo {author} {\bibfnamefont {L.}~\bibnamefont {Pang}},\ and\
  \bibinfo {author} {\bibfnamefont {Q.}~\bibnamefont {Wang}},\ }\bibfield
  {title} {\bibinfo {title} {{Elliptic flow of thermal dileptons in
  event-by-event hydrodynamic simulation}},\ }\href
  {https://doi.org/10.1103/PhysRevC.89.064902} {\bibfield  {journal} {\bibinfo
  {journal} {Phys. Rev. C}\ }\textbf {\bibinfo {volume} {89}},\ \bibinfo
  {pages} {064902} (\bibinfo {year} {2014})},\ \Eprint
  {https://arxiv.org/abs/1404.2663} {arXiv:1404.2663 [hep-ph]} \BibitemShut
  {NoStop}%
\bibitem [{\citenamefont {Vujanovic}\ \emph {et~al.}(2014)\citenamefont
  {Vujanovic}, \citenamefont {Young}, \citenamefont {Schenke}, \citenamefont
  {Rapp}, \citenamefont {Jeon},\ and\ \citenamefont
  {Gale}}]{Vujanovic:2013jpa}%
  \BibitemOpen
  \bibfield  {author} {\bibinfo {author} {\bibfnamefont {G.}~\bibnamefont
  {Vujanovic}}, \bibinfo {author} {\bibfnamefont {C.}~\bibnamefont {Young}},
  \bibinfo {author} {\bibfnamefont {B.}~\bibnamefont {Schenke}}, \bibinfo
  {author} {\bibfnamefont {R.}~\bibnamefont {Rapp}}, \bibinfo {author}
  {\bibfnamefont {S.}~\bibnamefont {Jeon}},\ and\ \bibinfo {author}
  {\bibfnamefont {C.}~\bibnamefont {Gale}},\ }\bibfield  {title} {\bibinfo
  {title} {{Dilepton emission in high-energy heavy-ion collisions with viscous
  hydrodynamics}},\ }\href {https://doi.org/10.1103/PhysRevC.89.034904}
  {\bibfield  {journal} {\bibinfo  {journal} {Phys. Rev. C}\ }\textbf {\bibinfo
  {volume} {89}},\ \bibinfo {pages} {034904} (\bibinfo {year} {2014})},\
  \Eprint {https://arxiv.org/abs/1312.0676} {arXiv:1312.0676 [nucl-th]}
  \BibitemShut {NoStop}%
\bibitem [{\citenamefont {Kasmaei}\ and\ \citenamefont
  {Strickland}(2019)}]{Kasmaei:2018oag}%
  \BibitemOpen
  \bibfield  {author} {\bibinfo {author} {\bibfnamefont {B.~S.}\ \bibnamefont
  {Kasmaei}}\ and\ \bibinfo {author} {\bibfnamefont {M.}~\bibnamefont
  {Strickland}},\ }\bibfield  {title} {\bibinfo {title} {{Dilepton production
  and elliptic flow from an anisotropic quark-gluon plasma}},\ }\href
  {https://doi.org/10.1103/PhysRevD.99.034015} {\bibfield  {journal} {\bibinfo
  {journal} {Phys. Rev. D}\ }\textbf {\bibinfo {volume} {99}},\ \bibinfo
  {pages} {034015} (\bibinfo {year} {2019})},\ \Eprint
  {https://arxiv.org/abs/1811.07486} {arXiv:1811.07486 [hep-ph]} \BibitemShut
  {NoStop}%
\bibitem [{\citenamefont {Chatterjee}\ \emph {et~al.}(2007)\citenamefont
  {Chatterjee}, \citenamefont {Srivastava}, \citenamefont {Heinz},\ and\
  \citenamefont {Gale}}]{Chatterjee:2007xk}%
  \BibitemOpen
  \bibfield  {author} {\bibinfo {author} {\bibfnamefont {R.}~\bibnamefont
  {Chatterjee}}, \bibinfo {author} {\bibfnamefont {D.~K.}\ \bibnamefont
  {Srivastava}}, \bibinfo {author} {\bibfnamefont {U.~W.}\ \bibnamefont
  {Heinz}},\ and\ \bibinfo {author} {\bibfnamefont {C.}~\bibnamefont {Gale}},\
  }\bibfield  {title} {\bibinfo {title} {{Elliptic flow of thermal dileptons in
  relativistic nuclear collisions}},\ }\href
  {https://doi.org/10.1103/PhysRevC.75.054909} {\bibfield  {journal} {\bibinfo
  {journal} {Phys. Rev. C}\ }\textbf {\bibinfo {volume} {75}},\ \bibinfo
  {pages} {054909} (\bibinfo {year} {2007})},\ \Eprint
  {https://arxiv.org/abs/nucl-th/0702039} {arXiv:nucl-th/0702039} \BibitemShut
  {NoStop}%
\bibitem [{\citenamefont {Adamczyk}\ \emph {et~al.}(2014)\citenamefont
  {Adamczyk} \emph {et~al.}}]{STAR:2014aok}%
  \BibitemOpen
  \bibfield  {author} {\bibinfo {author} {\bibfnamefont {L.}~\bibnamefont
  {Adamczyk}} \emph {et~al.} (\bibinfo {collaboration} {STAR}),\ }\bibfield
  {title} {\bibinfo {title} {{Dielectron azimuthal anisotropy at mid-rapidity
  in Au + Au collisions at $\sqrt{s_{_{NN}}} = 200$ GeV}},\ }\href
  {https://doi.org/10.1103/PhysRevC.90.064904} {\bibfield  {journal} {\bibinfo
  {journal} {Phys. Rev. C}\ }\textbf {\bibinfo {volume} {90}},\ \bibinfo
  {pages} {064904} (\bibinfo {year} {2014})},\ \Eprint
  {https://arxiv.org/abs/1402.1791} {arXiv:1402.1791 [nucl-ex]} \BibitemShut
  {NoStop}%
\bibitem [{\citenamefont {Yamamoto}\ and\ \citenamefont
  {Hirono}(2013)}]{Yamamoto:2013zwa}%
  \BibitemOpen
  \bibfield  {author} {\bibinfo {author} {\bibfnamefont {A.}~\bibnamefont
  {Yamamoto}}\ and\ \bibinfo {author} {\bibfnamefont {Y.}~\bibnamefont
  {Hirono}},\ }\bibfield  {title} {\bibinfo {title} {{Lattice QCD in rotating
  frames}},\ }\href {https://doi.org/10.1103/PhysRevLett.111.081601} {\bibfield
   {journal} {\bibinfo  {journal} {Phys. Rev. Lett.}\ }\textbf {\bibinfo
  {volume} {111}},\ \bibinfo {pages} {081601} (\bibinfo {year} {2013})},\
  \Eprint {https://arxiv.org/abs/1303.6292} {arXiv:1303.6292 [hep-lat]}
  \BibitemShut {NoStop}%
\bibitem [{\citenamefont {Deng}\ and\ \citenamefont
  {Huang}(2016)}]{Deng:2016gyh}%
  \BibitemOpen
  \bibfield  {author} {\bibinfo {author} {\bibfnamefont {W.-T.}\ \bibnamefont
  {Deng}}\ and\ \bibinfo {author} {\bibfnamefont {X.-G.}\ \bibnamefont
  {Huang}},\ }\bibfield  {title} {\bibinfo {title} {{Vorticity in Heavy-Ion
  Collisions}},\ }\href {https://doi.org/10.1103/PhysRevC.93.064907} {\bibfield
   {journal} {\bibinfo  {journal} {Phys. Rev. C}\ }\textbf {\bibinfo {volume}
  {93}},\ \bibinfo {pages} {064907} (\bibinfo {year} {2016})},\ \Eprint
  {https://arxiv.org/abs/1603.06117} {arXiv:1603.06117 [nucl-th]} \BibitemShut
  {NoStop}%
\bibitem [{\citenamefont {Wei}\ \emph {et~al.}(2020)\citenamefont {Wei},
  \citenamefont {Jiang},\ and\ \citenamefont {Huang}}]{Wei:2020xfd}%
  \BibitemOpen
  \bibfield  {author} {\bibinfo {author} {\bibfnamefont {M.}~\bibnamefont
  {Wei}}, \bibinfo {author} {\bibfnamefont {Y.}~\bibnamefont {Jiang}},\ and\
  \bibinfo {author} {\bibfnamefont {M.}~\bibnamefont {Huang}},\ }\bibfield
  {title} {\bibinfo {title} {{Mass splitting of vector meson and spontaneous
  spin polarization under rotation}},\ }\href@noop {} {\  (\bibinfo {year}
  {2020})},\ \Eprint {https://arxiv.org/abs/2011.10987} {arXiv:2011.10987
  [hep-ph]} \BibitemShut {NoStop}%
\bibitem [{\citenamefont {Voloshin}\ and\ \citenamefont
  {Zhang}(1996)}]{Voloshin:1994mz}%
  \BibitemOpen
  \bibfield  {author} {\bibinfo {author} {\bibfnamefont {S.}~\bibnamefont
  {Voloshin}}\ and\ \bibinfo {author} {\bibfnamefont {Y.}~\bibnamefont
  {Zhang}},\ }\bibfield  {title} {\bibinfo {title} {{Flow study in relativistic
  nuclear collisions by Fourier expansion of Azimuthal particle
  distributions}},\ }\href {https://doi.org/10.1007/s002880050141} {\bibfield
  {journal} {\bibinfo  {journal} {Z. Phys. C}\ }\textbf {\bibinfo {volume}
  {70}},\ \bibinfo {pages} {665} (\bibinfo {year} {1996})},\ \Eprint
  {https://arxiv.org/abs/hep-ph/9407282} {arXiv:hep-ph/9407282} \BibitemShut
  {NoStop}%
\bibitem [{Note1()}]{Note1}%
  \BibitemOpen
  \bibinfo {note} {The same limit taken in reverse order is known as isotropic
  limit.}\BibitemShut {Stop}%
\end{thebibliography}%

\end{document}